\documentclass[a4paper,openright,11pt]{article}
\usepackage{fancyhdr}
\usepackage{graphicx}
\usepackage{amsfonts}
\usepackage{amsmath, amsthm, amssymb}
\usepackage[utf8]{inputenc}
\usepackage{dsfont}
\usepackage[english]{babel}
\usepackage{cite}
\usepackage[hyperindex=true,breaklinks=true,colorlinks=true,linkcolor=black,citecolor=blue]{hyperref}

\title{Bilayer graphene coherent states}
\author{David J. Fern\'andez C.\footnote{david@fis.cinvestav.mx} \,\ and Dennis I. Mart\'inez-Moreno\footnote{dmartinez@fis.cinvestav.mx}\\
  \small Physics Department, Cinvestav, P.O. Box 14-740, 07000 Mexico City, Mexico \\
  \date{July 5, 2020}
}

\begin{document}

\maketitle

\begin{abstract}
In this paper we consider the interaction of electrons in bilayer graphene with a constant homogeneous magnetic field which is orthogonal to the bilayer surface. Departing from the energy eigenstates of the effective Hamiltonian, the corresponding coherent states will be constructed. For doing this, first we will determine appropriate creation and annihilation operators in order to subsequently derive the coherent states as eigenstates of the annihilation operator with complex eigenvalue. Then, we will calculate some physical quantities, as the Heisenberg uncertainty relation, the probabilities and current density as well as the mean energy value. Finally, we will explore the time evolution for these states and we will compare it with the corresponding evolution for monolayer graphene coherent states.
\end{abstract}

\section{Introduction}

Carbon is the basis of all organic chemistry; due to the flexibility of their bonds, carbon-based systems show different structures with a wide variety of physical properties \cite{castro}. Graphene consists of a monolayer of carbon atoms arranged in a hexagonal crystal lattice with a distance of 1.42\r{A} between nearest neighbour atoms. The discovery of this material and the interesting electronic properties induced by the low energy excitations have attracted attention of the scientific community. In particular, the possibility arises that certain aspects of relativistic quantum mechanics can be tested, like the Klein paradox or the anomalous Landau-Hall effect  \cite{castro,hall,kuru,NBOT17,CCJ20}.
\\

The electronic structure of graphene is typically studied using the \textit{tight-binding} model. Assuming that only next nearest neighbour hopping processes in the low-energy limit take place (close to the Dirac points), the equation ruling the electrons in graphene with applied static magnetic fields is the  Dirac-Weyl equation \cite{castro,kuru, mf14}. This equation can be solved for a constant homogeneous magnetic field, leading to the Landau levels of grapehene and the corresponding eigenfunctions, which constitutes the standard quantum mechanical framework \cite{kuru}\footnote{Let us note that the energy eigenfunctions and eigenvalues for graphene and some of its allotropes have been as well determined for time-independent magnetic fields which are not necessarily homogeneous \cite{kuru,MTM11,JKNT13,JKN14,mf14,CHRV18,ju19,cf20,PLLR20,FGO20} (see also \cite{IN18}).}. However, an alternative approach exists, in which the system is addressed through coherent states. This approach supplies information which is supplementary to that obtained by the standard method, and it has been recently implemented for monolayer graphene \cite{erick}.
\\

Bilayer graphene in a static magnetic field can be also studied through the \textit{tight-binding} model. Assuming that the electrons can hop between the next-nearest neighbours in the same layer or between layers via a perpendicular hopping parameter $t_{\bot}$, an equation with an effective Hamiltonian different from the monolayer one is obtained \cite{kats}. Once again, in the standard approach this problem can be solved exactly for a constant homogeneous magnetic field, leading to the Landau levels and the associated eigenfunctions for bilayer graphene, which are different from the monolayer ones.
\\

With this in mind, it seems natural to explore the possibility of generating the bilayer graphene coherent states (BGCS), thus addressing the alternative approach already implemented for monolayer graphene. This is the main goal of this paper, which is organized as follows. In section 2 some general aspects of bilayer graphene are discussed. Through the \textit{tight binding} model and some simplifying assumptions, the effective Hamiltonian in the low-energy approximation is obtained. The Landau levels for bilayer graphene in a magnetic field orthogonal to its surface are also determined. In section 3 the coherent states will be generated as eigenstates of the annihilation operator with complex eigenvalue, and several physical quantities for these states will be calculated. In section 4 the time evolution for the coherent states of bilayer and monolayer graphene are studied. Finally, our conclusions will be presented in section 5.

\section{General aspects of bilayer graphene}
\begin{figure}[t]
\begin{center} 
  \includegraphics[width=10cm]{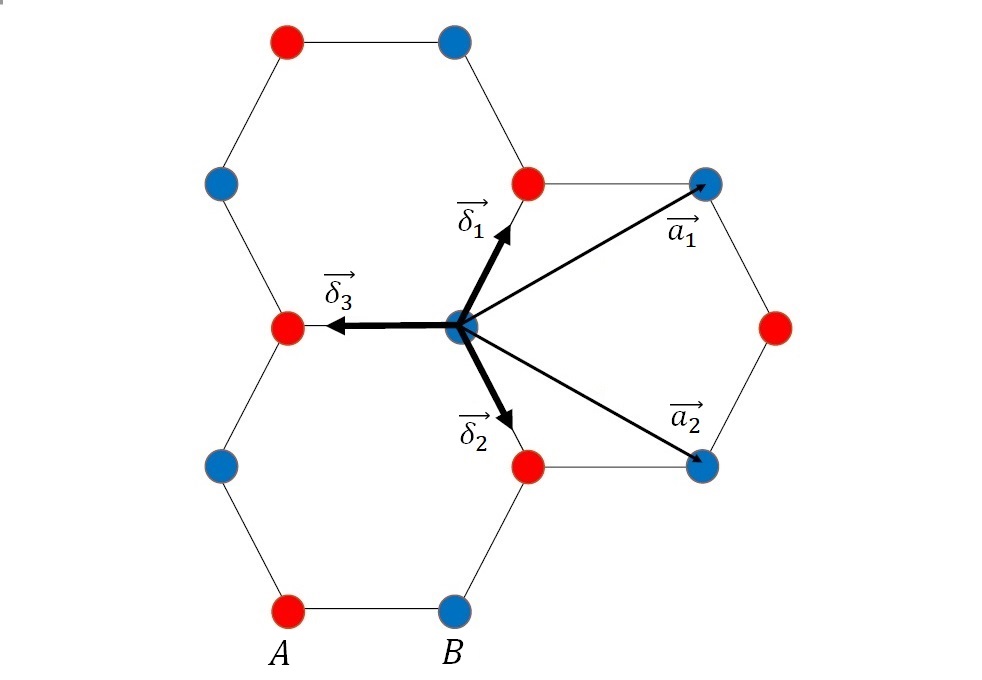}
   \caption{The lattice structure of monolayer graphene is made out of the two triangular sub-lattices A (red points) and B (blue points); $\vec{a}_{i}$ are the lattice vectors and $\vec{\delta}_{i}$ the nearest neighbour vectors.}
  \label{fig.grafeno}
  \end{center}  
\end{figure}

Graphene is formed by carbon atoms arranged in a honeycomb hexagonal crystal lattice, with two atoms per unit cell belonging to sub-lattices A and B. Each atom of sub-lattice A is surrounded by three atoms of sub-lattice B and vice versa, as it is shown in Figure \ref{fig.grafeno}.
\\

\begin{figure}[t]
\begin{center} 
  \includegraphics[width=13.5cm]{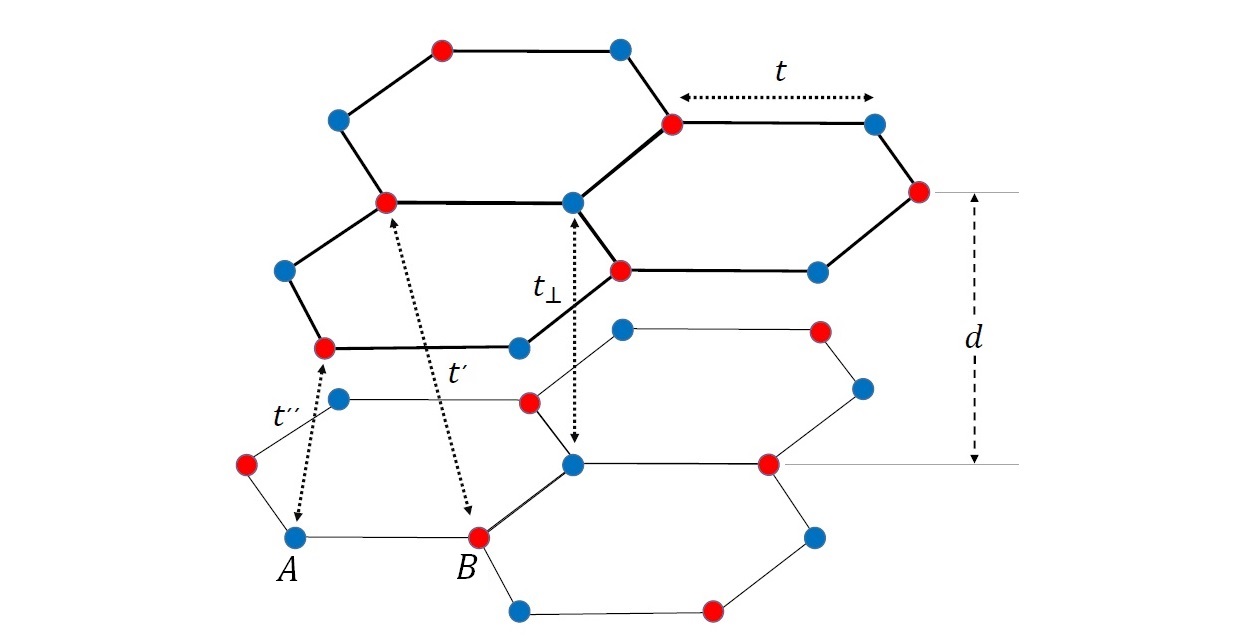}
   \caption{Lattice structure of bilayer graphene. The blue points correspond to the sub-lattices A and the red points to the sub-lattices B.}
     \label{fig.bicapa}
  \end{center}  
\end{figure}

We will consider here bilayer graphene, which is composed by two monolayers of carbon atoms and can be obtained by exfoliation of graphite. Its electronic structure can be studied in the framework of the \textit{tight-binding} model \cite{kats, MCMK2013}, while its crystal structure is shown in Figure \ref{fig.bicapa}. In bilayer graphene, the second layer of carbon atoms is rotated $60^{o}$ with respect to the first layer. Moreover, the sub-lattices A of the two layers lie exactly on top of one another. The hopping parameters are: $t\approx$ 2.8 eV is the hopping energy in the plane, between atoms of the same layer; 
$t_{\bot} \approx$ 0.4 eV is the hopping energy between the atom $A_{1}$ and the atom $A_{2}$; $t'\approx$ 0.3 eV is the hopping energy between the atom $B_{1}$ and the atom $B_{2}$; $t''\approx-$ 0.04 eV is the hopping energy between the atom $A_{1}(A_{2})$ and the atom $B_{2}(B_{1})$ \cite{castro}. The distance between the two graphene layers (inter-planar spacing) is $d\approx$ 3.35\r{A}.
\\

Let us consider the case where there are no hopping between the sublattices B of the two layers ($t'=0$) and  between crossed sublattices ($t''=0$). The model which takes into account these considerations \cite{kats} is characterized by the Hamiltonian 

\begin{equation}
H(\vec{k})=
\left(\begin{array}{cccc}
0 & tS(\vec{k}) & t_{\bot} & 0 \\ tS^{*}(\vec{k}) & 0 & 0 & 0 \\ t_{\bot} & 0 & 0 & tS^{*}(\vec{k}) \\ 0 & 0 & tS(\vec{k}) & 0 \\
\end{array}
\right), \quad \label{eq.ham1} 
\end{equation}
where
\begin{eqnarray*}
S(\vec{k})=\sum_{\vec{\delta}}e^{i\vec{k}\cdot\vec{\delta}}=2\exp\left( \dfrac{ik_{x}a}{2}\right) \cos\left( \dfrac{\sqrt{3}k_{y}a}{2}\right) +\exp\left( -ik_{x}a\right), 
\end{eqnarray*}
with $\vec{k}$ being the wave vector.
\\

In the low-energy approximation ($\vert E\vert \ll t_{\bot}$) the two important eigenvalues of the Hamiltonian of Eq. (\ref{eq.ham1}) are
\begin{equation}
E_{1,2}(\vec{k})\approx \pm \dfrac{t^{2}|S(\vec{k})|^{2}}{t_{\bot}}\approx \pm \dfrac{\hbar^{2}q^{2}}{2m^{*}}, \label{eq.eigen}
\end{equation}
where the effective mass is $m^{*}=|t_{\bot}|/(2v_{F}^{2})\approx$ 0.054$m_{e}$, with $m_{e}$ being the electron mass and $v_{F}$ the Fermi velocity in graphene. The other two energies $E_{3,4}(\vec{k})$ are separated by a gap of size $2|t_{\bot}|$, thus they are irrelevant for low-energies. \\
 
The replacement of $\hbar q_{x}$ and $\hbar q_{y}$ by the operators $\hat{p}_{x}=-i \hbar\frac{\partial}{\partial x}$ and $\hat{p}_{y}=-i \hbar\frac{\partial}{\partial y}$ respectively, leads to the following effective Hamiltonian for bilayer graphene
\begin{equation}
\hat{H}= \frac{1}{2m^*}
\left(\begin{array}{cc}
0 & (\hat{p}_{x} - i\hat{p}_{y})^2 \\ (\hat{p}_{x} + i\hat{p}_{y})^2 & 0\\
\end{array}
\right). \quad \label{eq.ham2}
\end{equation}

\subsection{Landau levels for bilayer graphene}
Let us consider now the bilayer graphene placed in a constant homogeneous magnetic field which is orthogonal to
the material surface (the \textit{x-y} plane). The interaction of the electrons with such field is described by the Hamiltonian of Eq. (\ref{eq.ham2}) where the momentum operator $\hat{\vec{p}}$ is replaced by $\hat{\vec{p}}+\dfrac{e}{c}\hat{\vec{A}}$ , according to the minimal coupling rule \cite{kuru, kats}, namely, 

\begin{equation}
\hat{H}= \frac{1}{2m^*}
\left(\begin{array}{cc}
0 & \left[ (\hat{p}_{x}+\dfrac{e}{c}\hat{A}_{x}) - i(\hat{p}_{y}+\dfrac{e}{c}\hat{A}_{y})\right] ^2 \\ \left[ (\hat{p}_{x}+\dfrac{e}{c}\hat{A}_{x}) + i(\hat{p}_{y}+\dfrac{e}{c}\hat{A}_{y})\right] ^2 & 0\\
\end{array}
\right). \quad \label{eq.ham3}
\end{equation}
In the Landau gauge the vector potential can be chosen as $\hat{\vec{A}}=\hat{A}(x)\hat{e}_{y}$ so that $\hat{\vec{B}}=\nabla\times\hat{\vec{A}}=\hat{B}(x)\hat{e}_{z}$. Additionally, in order to get a constant magnetic field orthogonal to the surface, along \textit{z} direction $\hat{\vec{B}}=B\hat{e}_{z}$, the vector potential is selected as $\hat{\vec{A}}=Bx\hat{e}_{y}$.
\\   

Let us define the operators (see \cite{kats})
\begin{equation}
\hat{\vec{\pi}}:=\hat{\vec{p}}+\frac{e}{c}\hat{\vec{A}}, \qquad \hat{\vec{p}}=-i\hbar\nabla, \label{eq.momento}
\end{equation}
such that
\begin{eqnarray*}
[\hat{\pi}_{x},\hat{\pi}_{y}]=-\dfrac{i\hbar e}{c}B \, \hat{1}.
\end{eqnarray*}

The bosonic operators $\hat{b}^{-}$ and $\hat{b}^{+}$ are introduced now as follows
\begin{equation}
\hat{\pi}_{-}:=\hat{\pi}_{x}-i\hat{\pi}_y=\sqrt{\frac{2e\hbar B}{c}} \, \hat{b}^{-}, \label{eq.bmenos}
\end{equation}
\begin{equation}
\hat{\pi}_{+}:= \hat{\pi}_{x}+i\hat{\pi}_y=\sqrt{\frac{2e\hbar B}{c}} \, \hat{b}^{+}, \label{eq.bmas}
\end{equation}
such that $[\hat{b}^{-},\hat{b}^{+}]=\hat{1}$. Thus, the effective Hamiltonian becomes

\begin{equation}
\hat{H}= \dfrac{1}{2m^{*}}
\left(\begin{array}{cc}
0 & \hat{\pi}_{-}^2 \\ \hat{\pi}_{+}^2 & 0\\
\end{array}
\right)=
\hbar \omega_{c}^{*}
\left(\begin{array}{cc}
0 & \hat{b}^-{}^2 \\ \hat{b}^+{}^2 & 0\\
\end{array}
\right), \,\,\,\ \omega_{c}^{*}=\frac{e B}{m^{*} c}, \label{eq.ham4}
\end{equation}
where $\omega_{c}^{*}$ is the cyclotron frequency for non-relativistic electrons with effective mass  $m^{*}$.\\

In order to find the stationary states, the time independent Schr\"odinger equation
\begin{equation}
\hat{H}\Psi(x,y)=E\Psi(x,y), \label{eq.sch}
\end{equation}
must be solved. Taking into account the translational invariance along $y$ direction, the two- component spinor is proposed as
\begin{equation}
\Psi(x,y)=\exp(iky)
\left(\begin{array}{c}
\psi^{+}(x) \\ \psi^{-}(x) \\
\end{array} 
\right), \label{eq.spinor}
\end{equation} 
where $k$ is the wave number in \textit{y}-direction and $\psi^{\pm}(x)$ describe the electron amplitude. Then, Eq. (\ref{eq.sch}) yields the next two equations:

\begin{equation}
(\hbar \omega_{c}^{*})(\hat{b}^{-})^2 \psi^{-}(x)=E\psi^{+}(x), \qquad
(\hbar \omega_{c}^{*})(\hat{b}^{+})^2 \psi^{+}(x)=E\psi^{-}(x). \label{eq.sol1}
\end{equation}\\
In order to decouple this system, let us calculate $\hat{H}^{2}\Psi(x,y)=E^{2}\Psi(x,y)$ so that

\begin{equation}
(\hbar \omega_{c}^{*})^{2}(\hat{b}^{-})^2(\hat{b}^{+})^2 \psi^{+}(x)=E^{2}\psi^{+}(x), \qquad
(\hbar \omega_{c}^{*})^{2}(\hat{b}^{+})^2(\hat{b}^{-})^2 \psi^{-}(x)=E^{2}\psi^{-}(x). \label{eq.sol2}
\end{equation}\\
It is well known that the harmonic oscillator Hamiltonian satisfies:
\begin{eqnarray} 
\hat{H}_{HO}&=&\hat{b}^{+} \hat{b}^{-}+ \dfrac{1}{2}, \nonumber \\ 
\hat{H}_{HO}\,\ \psi_{n}(x) &=&\left( n+\dfrac{1}{2}\right) \psi_{n}(x), \nonumber \\ 
\left[ \hat{b}^{-}, \hat{b}^{+}\right]&=&\hat{1}. \label{eq.osc}
\end{eqnarray}

If in Eq. (\ref{eq.sol2}) we take
\begin{eqnarray}
\psi^{+}(x):= \psi_{n-2}(x), \,\ \,\  \psi^{-}(x):= \psi_{n}(x), \label{eq.soluciones}
\end{eqnarray}
with $\psi_{n}(x)$ being the eigenfunctions of the harmonic oscillator Hamiltonian, it is obtained that

\begin{equation}
E^{2}=(\hbar \omega_{c}^{*})^{2}n(n-1).  \nonumber
\end{equation}

Thus, the eigenvalues of the Hamiltonian in Eq. (\ref{eq.ham4}) become
\begin{equation}
E_n^{\pm}=\pm \hbar \omega_{c}^{*} \sqrt{n(n-1)}, \label{eq.eigenvalues}
\end{equation}
where the plus (minus) sign characterizes the energy electrons (holes), while the normalized eigenstates are

\begin{center}
$\Psi_{0}(x,y)= \exp(iky)
\left(\begin{array}{c}
0 \\ \psi_{0}(x) \\
\end{array} 
\right),\qquad 
\Psi_{1}(x,y)= \exp(iky)
\left(\begin{array}{c}
0 \\ \psi_{1}(x) \\
\end{array} 
\right),$
\end{center}

\begin{equation}
\Psi_{n}(x,y)= \frac{\exp(iky)}{\sqrt{2}}
\left(\begin{array}{c}
\psi_{n-2}(x) \\ \psi_{n}(x) \\
\end{array} 
\right), \,\ \,\ n=2,3,\ldots  \label{eq.spinores} 
\end{equation}
with (see \cite{FGO20})
\begin{eqnarray}
\psi_{n}(x)=\sqrt{\dfrac{1}{2^{n} n!}\left(\dfrac{\omega}{2\pi} \right)^{1/2} }
H_{n}\left[\sqrt{\dfrac{\omega}{2}}\left(x+\dfrac{2k}{\omega} \right)  \right]
\exp\left( -\dfrac{\omega}{4}\left(x+\dfrac{2k}{\omega}  \right)^{2} \right), \,\ \omega=\dfrac{2m^{*}}{\hbar}\omega_{c}^{*}. \label{eq.solosc}
\end{eqnarray}

We should note that $E_{0}^{\pm}=E_{1}^{\pm}=0$, i.e., the ground state energy has a fourfold degeneracy (a double degeneracy due to electrons and the same due the holes). On the other hand, the levels $E_{n}^{\pm}$ for $n\geq2$ are non degenerate.

\section{Coherent states}
We will follow a procedure similar to the one used for monolayer graphene \cite{erick} in order to generate the bilayer graphene coherent states. Thus, we need to identify first the appropriate annihilation and creation operators, which will be done next.

\subsection{Annihilation and creation operators}

The annihilation operator $\hat{A}^{-}$ for the Hamiltonian in Eq. (\ref{eq.ham4}) is proposed as follows
\\
\begin{equation}
\hat{A}^{-}:=
\left(\begin{array}{cc}
f_{3}(\hat{N}) \hat{a}^{-} & 0 \\ 0 & f(\hat{N} + \hat{1}) \hat{a}^{-} \\
\end{array}
\right),  \label{eq.anniop}
\end{equation}
where the operators $\hat{a}^{\pm}$ and $\hat{N}$ are given by\\
\begin{eqnarray*}
\hat{a}^{\pm}= \frac{1}{\sqrt{2}} \left( \xi \mp \frac{d}{d\xi}\right) , \qquad \hat{N}=\hat{a}^{+}\hat{a}^{-},
\end{eqnarray*}
$\xi=\sqrt{\dfrac{\omega}{2}}\left(x+\dfrac{2k}{\omega}\right)$, and $f$, $f_{3}$ are two auxiliary real functions. The operators $\hat{a}^{\pm}$ and $\hat{b}^{\pm}$ are related as follows
\begin{eqnarray*}
i\, \hat{a}^{+}=\hat{b}^{+}, \,\ \,\ i\, \hat{a}^{-}=-\hat{b}^{-}. 
\end{eqnarray*}

Note that
\begin{equation}
\hat{A}^{-} \Psi_{n}(x,y)=\dfrac{\exp(iky)}{\sqrt{2}} \left(\begin{array}{c}
\sqrt{n-2}\,f_{3}(n-3)\, \psi_{n-3}(x) \\ \sqrt{n}\,f(n)\,\psi_{n-1}(x) \\
\end{array} 
\right). \label{eq.anniop1}
\end{equation}
In order to ensure that $\hat{A}^{-} \Psi_{n}(x,y)=c_{n}\Psi_{n-1}(x,y)$, the functions $f$, $f_{3}$ must fulfill 
\begin{equation}
\sqrt{n-2} f_{3}(n-3)=\sqrt{n} f(n)\equiv c_{n}. \label{eq.funciones}
\end{equation}
Thus, the explicit expression for the annihilation operator $\hat{A}^{-}$ becomes
\begin{equation}
\hat{A}^{-}=
\left(\begin{array}{cc}
\frac{\sqrt{\hat{N}+ \hat{3}}}{\sqrt{\hat{N} + \hat{1}}} f(\hat{N}+\hat{3}) \hat{a}^{-} & 0 \\ 0 & f(\hat{N} + \hat{1}) \hat{a}^{-} \\
\end{array}
\right). \label{eq.anniop2}
\end{equation}

Finally, the creation operator $\hat{A}^{+}$ is just the Hermitian conjugate of $\hat{A}^{-}$, i.e.,

\begin{equation}
\hat{A}^{+}=
\left(\begin{array}{cc}
\hat{a}^{+} \frac{\sqrt{\hat{N}+ \hat{3}}}{\sqrt{\hat{N} + \hat{1}}} f(\hat{N}+\hat{3}) & 0 \\ 0 & \hat{a}^{+} f(\hat{N} + \hat{1})  \\
\end{array}
\right), \quad \label{eq.creatop}
\end{equation}
such that
\begin{equation}
\begin{split}
\hat{A}^{+}\Psi_{0}(x,y)=& f(1)\Psi_{1}(x,y), \\
\hat{A}^{+}\Psi_{1}(x,y)=& \sqrt{2}f(2)\exp(iky)
\left(\begin{array}{c}
0 \\ \psi_{2}(x) 
\end{array}
\right), \\
\hat{A}^{+}\Psi_{n}(x,y)= &\sqrt{n+1}f(n+1)\Psi_{n+1}(x,y), \,\ \,\ n=2,3,\ldots \, 
\end{split} \label{eq.creatop1}
\end{equation}

\subsection{Coherent states as eigenstates of $\hat{A}^{-}$}
The coherent states $\Psi_{\alpha}(x,y)$, discovered for the first time by Schr\"odinger in 1926 \cite{sc} and rediscovered later by Klauder, Glauber and Sudarshan in the early 1960s \cite{glauber1, glauber2, glauber3}, are quantum states very close to classical states \cite{coherente}. They can be defined as eigenstates of the previous annihilation operator $\hat{A}^{-}$ with complex eigenvalue $\alpha$, namely, 

\begin{equation}
\hat{A}^{-}\Psi_{\alpha}(x,y)=\alpha\Psi_{\alpha}(x,y), \,\ \qquad  \,\  \alpha\in \mathbb{C}. \label{eq.coherent}
\end{equation}

By expressing now the states $\Psi_{\alpha}(x,y)$ as a linear combination of the eigenstates $\Psi_{n}(x,y)$,
\begin{equation}
\begin{split}
\Psi_{\alpha}(x,y)= \sum_{n = 0}^\infty a_{n}\Psi_{n}(x,y)=e^{iky}  
\left[ a_{0}\left(\begin{array}{c}
0 \\ \psi_{0}(x) 
\end{array}
\right) + 
a_{1}\left(\begin{array}{c}
0 \\ \psi_{1}(x) 
\end{array}
\right) + 
\sum_{n = 2}^\infty \frac{a_{n}}{\sqrt{2}}\left(\begin{array}{c}
\psi_{n-2}(x) \\ \psi_{n}(x) 
\end{array}
\right)\right],
\end{split} \label{eq.expansion}
\end{equation}
and substituting this expression into equation (\ref{eq.coherent}) we obtain a recurrence relationship for the coefficients $a_{n}$
\begin{equation}
\begin{split}
a_{1}=& \frac{\alpha a_{0}}{f(1)}, \\
a_{2}=& \frac{\alpha a_{1}}{f(2)}, \\
a_{n+1}=& \frac{\alpha}{\sqrt{n+1}f(n+1)}a_{n}, \,\ n=2,3, \dots \label{eq.coef}
\end{split}
\end{equation}

We can identify two different cases.
\subsubsection{Coherent states when $f(1)\neq0$}
Suppose that $f(n)\neq0$\, $\forall \, n\in \mathbb{N}$, so that the coefficients $a_{n}$ become
\begin{equation}
a_{n}=\frac{\sqrt{2}\alpha^{n}}{\sqrt{n!}\,[f(n)]!}a_{0}, \qquad n=2,3,\ldots \label{eq.coef1}
\end{equation}
where, for any function $q(s)$ such that $s\in \mathbb{N}$, the generalized factorial function is defined by 
$$
[q(s)]!:=\left\{
	       \begin{array}{ll}
		 1 \,\  \,\ \,\ \,\ \,\ \,\ \,\ \,\ \,\ \,\ \,\ \,\  \textrm {for}   & s=0, \\
		 q(1) \cdot\cdot\cdot q(s) \,\ \,\ \textrm{for} & s>0.
		 \end{array}
	     \right. 
$$
Thus, the coherent states in this case are
\begin{multline}
\Psi_{\alpha}(x,y)= \left[ 1 + \frac{|\alpha|^{2}}{f(1)^{2}} + 2  \sum_{n = 2}^\infty \frac{|\alpha|^{2n}}{n!\, ([f(n)]!)^{2}}\right] ^{- \frac{1}{2}} \\
 \times \left[ \Psi_{0}(x,y) + \frac{\alpha}{f(1)} \, \Psi_{1}(x,y) + \sqrt{2} \sum_{n = 2}^\infty \frac{\alpha^{n} }{\sqrt{n!}\,[f(n)]!} \,  \Psi_{n}(x,y)\right] . \label{eq.coherente1}
\end{multline}

Note that the states of Eq. (\ref{eq.coherente1}) were normalized using the  free coefficient $a_{0}$.

\subsubsection{Coherent states when $f(1)=0$}
If $f(1)=0$ it should happen that $a_{0}=0$. Once again, two different subcases arise.

\subsubsection*{A. Case with $f(2)\neq0$. }
If $f(2)\neq 0$ the free parameter becomes now $a_{1}$. The recurrence relationship for the coefficients $a_{n}$ leads to

\begin{equation}
a_{n+1}= \frac{\sqrt{2} \alpha^{n}}{\sqrt{(n+1)!}\,[g(n)]!} a_{1}, \,\ n=1,2, \dots \label{eq.coef2}
\end{equation}
\\
were $g(n):= f(n+1)$. Substituting this result in Eq. (\ref{eq.expansion}), the normalized coherent states are
\begin{equation}
\Psi_{\alpha}(x,y) = \left[  1 + 2\sum_{n = 1}^\infty \frac{|\alpha|^{2n}}{(n+1)!\, ([g(n]]!)^{2}}\right] ^{- \frac{1}{2}}\left[ \Psi_{1}(x,y) + \sqrt{2}\sum_{n = 1}^\infty \frac{\alpha^{n}}{\sqrt{(n+1)!}\,[g(n)]!}\, \Psi_{n+1}(x,y)\right]. \label{eq.coherente2}
\end{equation}

\subsubsection*{B. Case with $f(2)=0$.} 

On the other hand, if $f(2)=0$ it should happen that $a_{1}=0$, therefore $a_{2}$ is the free parameter now. By defining $h(n):= f(n+2) \,\ \forall \,\ n\in \mathbb{N}$, the normalized coherent states become now
\begin{equation}
\Psi_{\alpha}(x,y) = \left[ \sum_{n = 0}^\infty \frac{|\alpha|^{2n}}{(n+2)! \, ([h(n)]!)^{2}}\right] ^{- \frac{1}{2}}\sum_{n = 0}^\infty \frac{ \alpha^{n}}{\sqrt{(n+2)!}\,[h(n)]!}\, \Psi_{n+2}(x,y). \label{eq.coherente3}
\end{equation} 

Equations (\ref{eq.coherente1}, \ref{eq.coherente2}, \ref{eq.coherente3}) contain three different sets of bilayer graphene coherent states (BGCS), all of them depending on the particular choice of the function $f(n)$. Note that these BGCS look similar to the monolayer graphene coherent states (MGCS) derived in \cite{erick}. However, as we will see later on the BGCS evolve in time in a completely different way as the MGCS do (see Section 4).

\subsection{Mean values and Heisenberg uncertainty relation}
The Heisenberg uncertainty relation (HUR) has been useful for studying the standard coherent states, since it is one of the most important quantities available to analyse the possible classical behaviour of a given quantum state \cite{coherente}.
\\

In order to introduce the HUR, the dimensionless position and momentum operators in terms of $\hat{a}^{\pm}$ are required, i.e.,
\begin{eqnarray*}
\hat{q} = \frac{1}{\sqrt{2}}( \hat{a}^{+} + \hat{a}^{-}), \qquad  \hat{p} = \frac{i}{\sqrt{2}}( \hat{a}^{+} - \hat{a}^{-}).
\end{eqnarray*}
Then, the Heisenberg uncertainty relation for $\hat{q}$ and $\hat{p}$ in a coherent state $\Psi_{\alpha}$ is given by
\begin{equation}
(\sigma_{q})_{\alpha} (\sigma_{p})_{\alpha} \geq \frac{1}{2}, \label{eq.uncertainty}
\end{equation}
where the standard deviation for an arbitrary observable $\hat{S}$ is defined as follows
\[ \sigma_{S}:=\sqrt{ \langle \hat{S}^{2}\rangle - \langle \hat{S} \rangle^{2} }.\]

We are going to calculate next the HUR for the BGCS built in the previous section.

\subsubsection{Mean values when $f(1)\neq0$}
First of all, let us make the particular choice $f(\hat{N}):= \hat{1}$. Thus, Eq. (\ref{eq.coherente1}) reduces to
\begin{equation}
\Psi_{\alpha}(x,y) = \frac{1}{\sqrt{2e^{r^{2}} - r^{2} - 1}} \left[ \Psi_{0}(x,y) + \alpha \Psi_{1}(x,y) + \sqrt{2} \sum_{n=2}^\infty \frac{ \alpha^{n}}{\sqrt{n!}} \, \Psi_{n}(x,y) \right], \label{eq.coherente11}
\end{equation}
where $\alpha=r \exp(i\theta)$.
\\

\begin{figure}[t]
\begin{center} 
  \includegraphics[width=16cm, height=7.5cm]{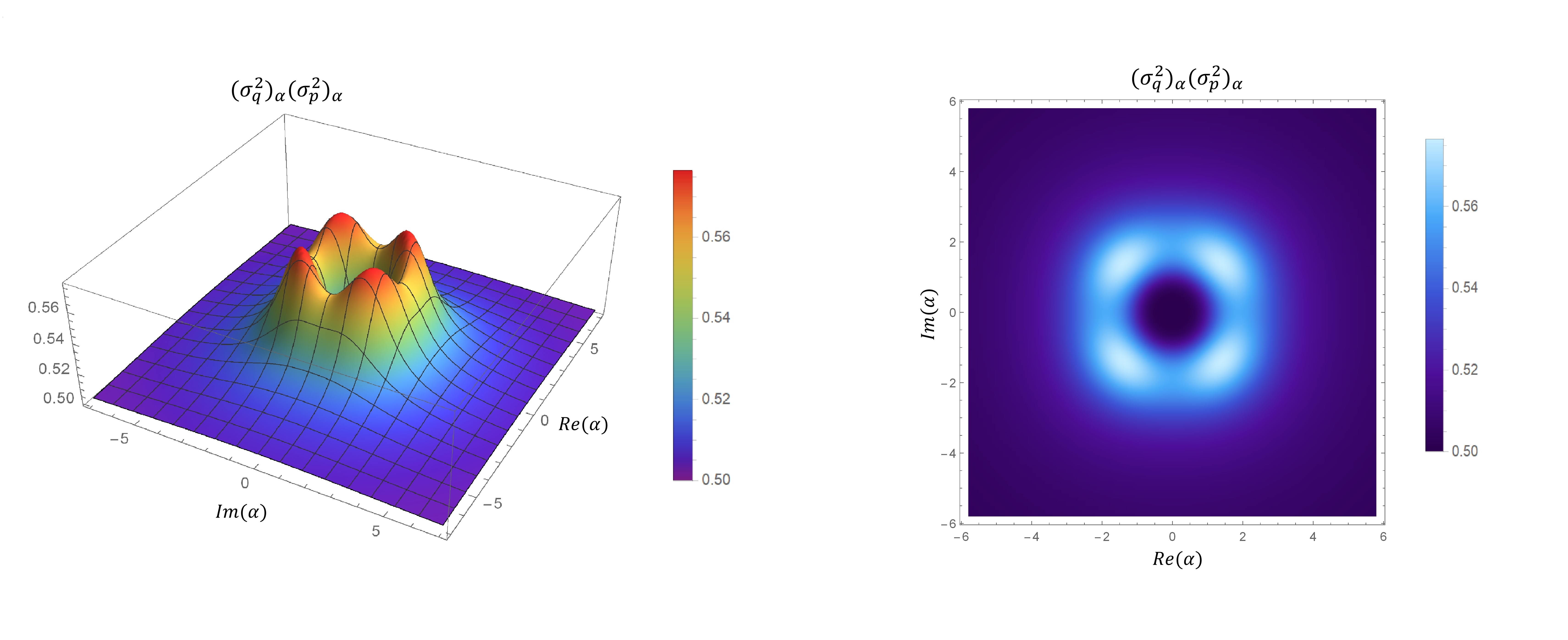}
   \caption{Heisenberg uncertainty relation $(\sigma_{q})_{\alpha} (\sigma_{p})_{\alpha}$ as function of $\alpha$ for the BGCS in the case that $f(n)=1$.}
  \label{fig.incertidumbre1}
  \end{center}  
\end{figure}

The mean values for $\hat{q}$, $\hat{p}$ and their squares become
\begin{subequations}
\begin{align}
\langle \hat{q}\rangle_{\alpha}&= \frac{\sqrt{2} \, \textrm{Re}(\alpha)}{2e^{r^{2}} - r^{2} - 1} \left[ \exp(r^{2}) + \sum_{n=2}^\infty \frac{\sqrt{n-1}\,r^{2n}}{\sqrt{n! \, (n+1)!}} \right], \\
\langle \hat{p}\rangle_{\alpha}&= \frac{\sqrt{2} \, \textrm{Im}(\alpha)}{2e^{r^{2}} - r^{2} - 1} \left[ \exp(r^{2}) + \sum_{n=2}^\infty \frac{\sqrt{n-1} \, r^{2n}}{\sqrt{n! \, (n+1)!)}} \right],\\
\langle \hat{q}^{2}\rangle_{\alpha}&= \frac{1}{4e^{r^{2}} - 2r^{2}- 2}
\left[ 1 + 3r^{2} + 2\sum_{n=2}^\infty \frac{(2n-1)r^{2n}}{n!} + 2\left( [\textrm{Re}(\alpha)]^{2} - [\textrm{Im}(\alpha)]^{2}\right)  \nonumber \right. \\ 
&\left. \,\ \,\ \,\ \,\ \,\ \,\ \,\ \,\ \,\ \,\ \,\ \,\ \,\ \,\ \,\ \,\ \,\ \,\ \,\ \,\ \,\ \,\ \,\ \,\ \,\ \,\ \,\ \,\  \times \left( \exp(r^{2}) + \sum_{n=2}^\infty \frac{r^{2n}}{\sqrt{(n-2)! \, (n+2)!}} \right)\right],\\
\langle \hat{p}^{2}\rangle_{\alpha}&= \frac{1}{4e^{r^{2}} - 2r^{2}- 2}
\left[ 1 + 3r^{2} + 2\sum_{n=2}^\infty \frac{(2n-1)r^{2n}}{n!}-2\left( [\textrm{Re}(\alpha)]^{2} - [\textrm{Im}(\alpha)]^{2}\right)  \nonumber \right. \\
&\left. \,\ \,\ \,\ \,\ \,\ \,\ \,\ \,\ \,\ \,\ \,\ \,\ \,\ \,\ \,\ \,\ \,\ \,\ \,\ \,\ \,\ \,\ \,\ \,\ \,\ \,\ \,\ \,\ \times \left( \exp(r^{2}) + \sum_{n=2}^\infty \frac{r^{2n}}{\sqrt{(n-2)! \, (n+2)!}} \right)\right]. \label{eq.momentos1}
\end{align}
\end{subequations}

From these the Heisenberg uncertainty relation $(\sigma_{q})_{\alpha} (\sigma_{p})_{\alpha}$ can be calculated (see a plot in Figure \ref{fig.incertidumbre1}). Note that  $(\sigma_{q})_{\alpha} (\sigma_{p})_{\alpha}\rightarrow1/2$ when $\alpha\rightarrow0$, which is due to the BGCS in Eq. (\ref{eq.coherente11}) tend to the eigenstate with minimum proper energy involved in the expansion, i.e., $\Psi_{0}(x,y)$. This is the lowest value that the Heisenberg uncertainty relation can have, and it coincides with the one obtained for the standard coherent states \cite{Gerry}.

\subsubsection{Mean values when $f(1)=0$}
\subsubsection*{A. Case with $f(2)\neq 0$.}
Let us choose now $f( \hat{N} + \hat{1}) := g( \hat{N}):= \frac{\sqrt{\hat{N}}}{\sqrt{ \hat{N}+ \hat{1}}}$, so that $f(n+1) \neq 0 \,\ \forall \,\ n=1,2,\dots$. Then, the coherent states of Eq. (\ref{eq.coherente2}) become

\begin{equation}
\Psi_{\alpha}(x,y) = \frac{1}{\sqrt{2e^{r^{2}}-1}} \left[ \Psi_{1}(x,y) + \sqrt{2} \sum_{n=1}^\infty \frac{\alpha^{n}}{\sqrt{n!}} \Psi_{n+1}(x,y) \right]. \label{eq.coherente22}  
\end{equation}

The mean values for $\hat{q}$, $\hat{p}$ and their squares in this state are given by

\begin{subequations}
\begin{align}
\langle \hat{q}\rangle_{\alpha} &= \frac{\sqrt{2} \, \textrm{Re} (\alpha)}{2e^{r^{2}}-1} \left[ \sum_{n=0}^\infty \frac{\sqrt{n+2} \, r^{2n}}{\sqrt{n! \,(n+1)!}} + \sum_{n=1}^\infty \frac{r^{2n}}{\sqrt{(n-1)! \, (n+1)!}} \right],\\
\langle \hat{p}\rangle_{\alpha} &= \frac{\sqrt{2} \, \textrm{Im} (\alpha)}{2e^{r^{2}}-1} \left[ \sum_{n=0}^\infty \frac{\sqrt{n+2} \, r^{2n}}{\sqrt{n! \, (n+1)!}} + \sum_{n=1}^\infty \frac{r^{2n}}{\sqrt{(n-1)! \, (n+1)!}} \right],\\
\langle \hat{q}^{2}\rangle_{\alpha} &= \frac{1}{4e^{r^{2}}-2} \left[  1+2\sum_{n=0}^\infty \frac{(2n+1)r^{2n}}{n!}+ 2\left( [\textrm{Re}(\alpha)]^{2} - [\textrm{Im}(\alpha)]^2 \right)   \nonumber \right. \\
&\left. \,\ \,\ \,\ \,\ \,\ \,\ \,\ \,\ \,\ \,\ \,\ \,\ \,\ \,\   \times \left(\sum_{n=1}^\infty \frac{\sqrt{n+1} \, r^{2n}}{\sqrt{(n-1)! \, (n+2)!}} +  \sum_{n=0}^\infty \frac{\sqrt{n+3} \, r^{2n}}{\sqrt{ n! \, (n+1)!}}   \right) \right],\\
\langle \hat{p}^{2}\rangle_{\alpha} &= \frac{1}{4e^{r^{2}}-2} \left[ 1+2\sum_{n=0}^\infty \frac{(2n+1)r^{2n}}{n!}- 2\left( [\textrm{Re}(\alpha)]^{2} - [\textrm{Im}(\alpha)]^2 \right)   \nonumber \right. \\
&\left. \,\ \,\ \,\ \,\ \,\ \,\ \,\ \,\ \,\ \,\ \,\ \,\ \,\ \,\  \times \left(\sum_{n=1}^\infty \frac{\sqrt{n+1} \, r^{2n}}{\sqrt{ (n-1)! \, (n+2)!}} +  \sum_{n=0}^\infty \frac{\sqrt{n+3} \, r^{2n}}{\sqrt{ n! \, (n+1)!}}   \right) \right]. \label{eq.momentos2}
\end{align}
\end{subequations}

\begin{figure}[t]
\begin{center} 
  \includegraphics[width=16cm, height=8.2cm]{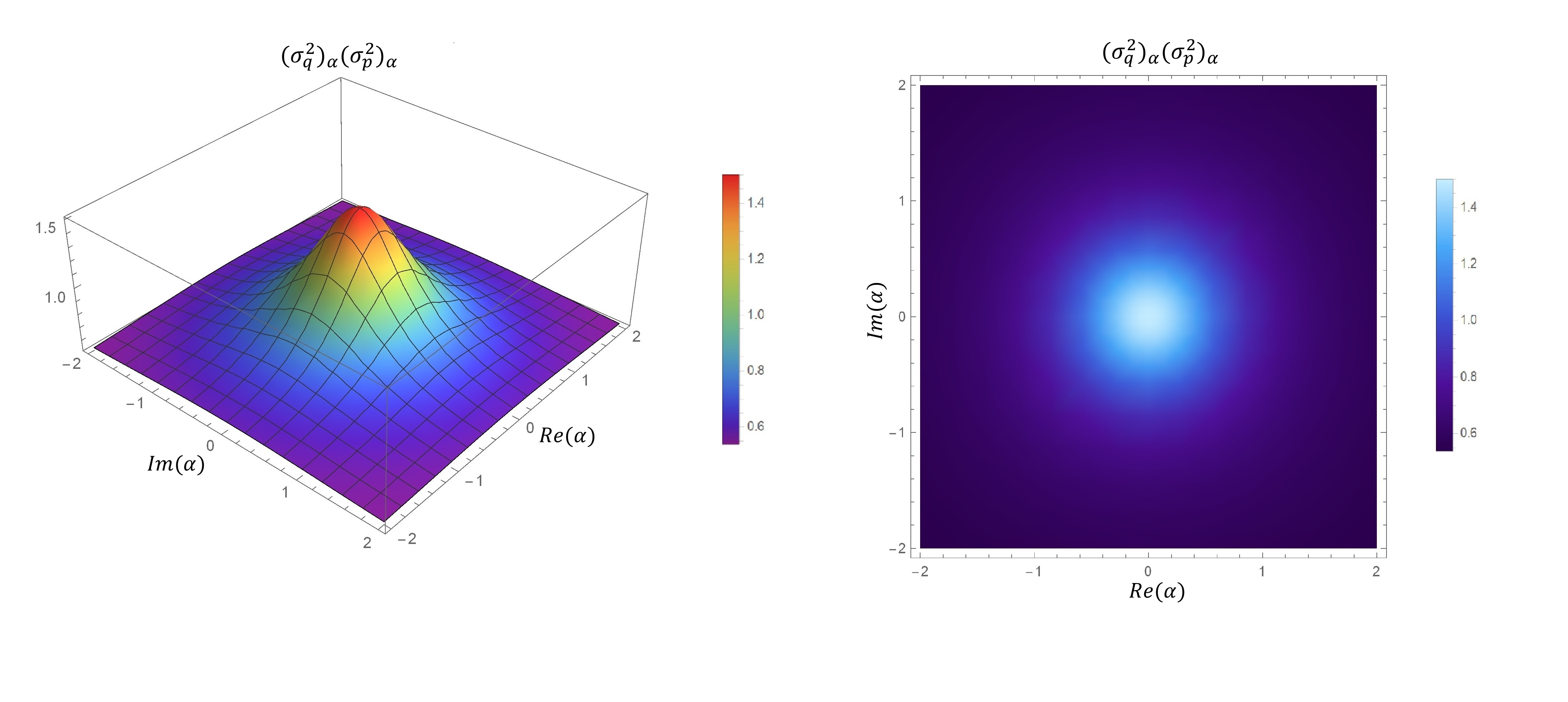}
   \caption{Heisenberg uncertainty relation $(\sigma_{q})_{\alpha} (\sigma_{p})_{\alpha}$ as function of $\alpha$ for the BGCS in the case that $f(n)=\sqrt{n-1}/\sqrt{n}$.}
  \label{fig.incertidumbre2}
  \end{center}  
\end{figure}

Note that $(\sigma_{q})_{\alpha} (\sigma_{p})_{\alpha}\rightarrow3/2$ when $\alpha\rightarrow0$ (see Figure \ref{fig.incertidumbre2}). This is so since the BGCS of Eq. (\ref{eq.coherente22}) tend to $\Psi_{1}(x,y)$ in this limit, and the Heisenberg uncertainty relation reaches now a maximum for $\alpha=0$.

\subsubsection*{B. Case with $f(2)=0$.}
Finally, let us consider that $f(\hat{N} + \hat{2}) = h(\hat{N}) = \frac{\hat{N} \sqrt{\hat{N} + \hat{1} }}{\sqrt{\hat{N} + \hat{2}}}$, thus the normalized coherent states of Eq. (\ref{eq.coherente3}) turn out to be
\begin{equation}
\Psi_{\alpha}(x,y) = \frac{1}{\sqrt{\,\ {_0}F_{2}(1,2;r^{2})}}\left[ \sum_{n=0}^\infty \frac{\alpha^{n}}{n! \sqrt{(n+1)!}} \Psi_ {n+2}(x,y) \right], \label{eq.coherente33}
\end{equation} 
were ${_p}F_{q}$ is the generalized hypergeometric function defined by

\begin{equation}
{_p}F_{q}(a_{1}, \ldots,a_{p},b_{1}\ldots,b_{q};x)=\dfrac{\Gamma(b_{1})\cdots\Gamma(b_{q})}{\Gamma(a_{1})\cdots\Gamma(a_{p})}\sum_{n=0}^\infty \dfrac{\Gamma(a_{1}+n)\cdots \Gamma(a_{p}+n)}{\Gamma(b_{1}+n)\cdots\Gamma(b_{q}+n)}\dfrac{x^{n}}{n!}.  \label{eq.hiper} 
\end{equation}
\\

The mean values for the position and momentum operators and their squares are now

\begin{subequations}
\begin{align}
\langle\hat{q}\rangle_{\alpha} &= \frac{1}{\sqrt{2}}\frac{\textrm{Re}(\alpha)}{{_0}F_{2}(1,2;r^{2})}\left[ \sum_{n=0}^\infty       \frac{r^{2n}}{(n+1)!\sqrt{(n+2)! \, (n!)^{3}}}  \nonumber \right. \\
&\left. \,\ \,\ \,\ \,\ \,\ \,\ \,\ \,\ \,\ \,\ \,\ \,\ \,\ \,\ \,\ \,\ \,\ \,\ \,\ \,\ \,\ \,\ \,\ \,\ \,\ \,\ \,\ \,\ \,\ \,\ \,\ \,\ \,\ \,\ \,\ \,\  +\sum_{n=0}^\infty \frac{\sqrt{n+3} \, r^{2n}}{n!\sqrt{(n+2)! \, [(n+1)!]^{3}}} \right], \\
\langle\hat{p}\rangle_{\alpha} &= \frac{1}{\sqrt{2}}\frac{\textrm{Im}(\alpha)}{{_0}F_{2}(1,2;r^{2})}\left[ \sum_{n=0}^\infty       \frac{r^{2n}}{(n+1)!\sqrt{(n+2)! \, (n!)^{3}}}  \nonumber \right. \\
&\left. \,\ \,\ \,\ \,\ \,\ \,\ \,\ \,\ \,\ \,\ \,\ \,\ \,\ \,\ \,\ \,\ \,\ \,\ \,\ \,\ \,\ \,\ \,\ \,\ \,\ \,\ \,\ \,\ \,\ \,\ \,\ \,\ \,\ \,\ \,\ \,\  +\sum_{n=0}^\infty \frac{\sqrt{n+3} \, r^{2n}}{n! \sqrt{(n+2)! \, [(n+1)!]^{3}}} \right],\\
\langle \hat{q}^{2} \rangle_{\alpha} &= \frac{1}{2\,{_0}F_{2}(1,2;r^{2})}\left[ \sum_{n=0}^\infty \frac{(2n+3)r^{2n}}{(n!)^{2} \, (n+1)!} + \left( [\textrm{Re(}\alpha)]^{2} - [\textrm{Im}(\alpha)]^2 \right) \nonumber \right. \\
&\left. \,\ \,\ \,\ \,\ \,\ \,\ \,\  \times \left(\sum_{n=0}^\infty \frac{\sqrt{n+2} \, r^{2n}}{(n+2)! \sqrt{(n+3)! \, (n!)^{3}}} + \sum_{n=0}^\infty \frac{\sqrt{n+4} \, r^{2n}}{ n!\sqrt{(n+1)! \, [(n+2)!]^{3}}}  \right)\right],\\
\langle \hat{p}^{2} \rangle_{\alpha} &= \frac{1}{2\,{_0}F_{2}(1,2;r^{2})}\left[ \sum_{n=0}^\infty \frac{(2n+3)r^{2n}}{(n!)^{2} \, (n+1)!} - \left( [\textrm{Re}(\alpha)]^{2} - [\textrm{Im}(\alpha)]^2 \right) \nonumber \right. \\
&\left. \,\ \,\ \,\ \,\ \,\ \,\ \,\ \times \left(\sum_{n=0}^\infty \frac{\sqrt{n+2} \, r^{2n}}{(n+2)! \sqrt{(n+3)! \, (n!)^{3}}} + \sum_{n=0}^\infty \frac{\sqrt{n+4} \, r^{2n}}{n!\sqrt{(n+1)! \, [(n+2)!]^{3}}}  \right)\right].  \label{eq.momentos3}
\end{align}
\end{subequations}

\begin{figure}[!]
\begin{center} 
  \includegraphics[width=16cm, height=8.5cm]{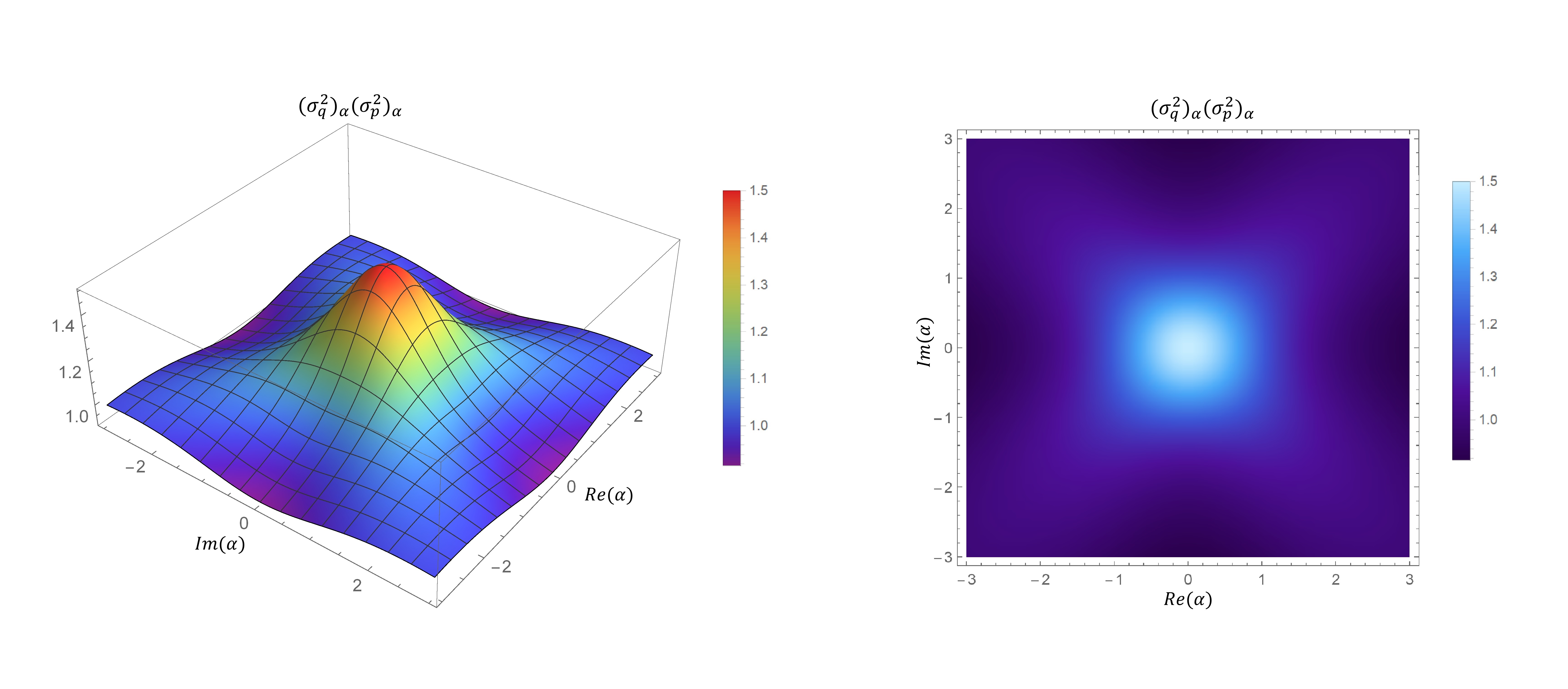}
    \caption{Heisenberg uncertainty relation $(\sigma_{q})_{\alpha} (\sigma_{p})_{\alpha}$ as function of $\alpha$ for the BGCS in the case that $f(n)=(n-2)\sqrt{n-1}/\sqrt{n}$.}
  \label{fig.incertidumbre3}
  \end{center}  
\end{figure}

Note that $(\sigma_{q})_{\alpha} (\sigma_{p})_{\alpha}\rightarrow3/2$ when $\alpha\rightarrow0$ (see Figure \ref{fig.incertidumbre3}). As before, this agrees with the fact that in this limit the BGCS of Eq. (\ref{eq.coherente33}) tend to the state with minimum energy involved in the lineal combination, i.e., $\Psi_{2}(x,y)$. Once again, the Heisenberg uncertainty relation reaches a maximum in this limit.

 \subsection{Probability density and probability current }
The probability density is defined by
\begin{equation}
\rho_{\alpha}(x) := \Psi^{\dagger}_{\alpha}(x,y) \Psi_{\alpha}(x,y). \label{eq.probabilidad}
\end{equation}

For the states given in Eq. (\ref{eq.spinores}) we will use the notation
\begin{equation}
\rho_{n,m}(x):= \psi_{n-2}(x) \psi_{m-2}(x) + \psi_{n}(x) \psi_{m}(x)=\rho_{m,n}(x), \qquad n=2,3,\dots \label{eq.rho}
\end{equation}
Note that for $n=0$ and $n=1$ we will have simply that
\begin{equation}
\begin{split}
\rho_{0}(x)=& \psi_{0}(x) \psi_{0}(x)=\mid\psi_{0}(x)\mid^{2}, \\
\rho_{1}(x)=& \psi_{1}(x) \psi_{1}(x)= \mid\psi_{1}(x)\mid^{2}.  \label{eq.rho12}
\end{split}
\end{equation}
\\
\\
\\

On the other hand, the probability current for electrons described by the Hamiltonian (\ref{eq.ham4}) takes the form \cite{corriente}

\begin{equation}
J_{\textit{l},\alpha}=\dfrac{\hbar}{m^{*}} \, \textrm{Im} \, \left[ \Psi_{\alpha}^{\dagger}(x,y)\, j_{\textit{l}} \, \, \Psi_{\alpha}(x,y)\right], \qquad \textit{l}=x, y, \label{eq.corriente}
\end{equation}
where

\begin{subequations}
\begin{align}
j_{x}=\sigma_{x}\partial_{x} + \sigma_{y}\partial_{y},\\
j_{y}=\sigma_{y}\partial_{x} - \sigma_{x}\partial_{y}, \label{eq.corrientecomp}
\end{align}
\end{subequations}
with $\sigma_{i}$ being the Pauli matrices.
\\
 
These quantities are time independent for stationary states, and they are also independent of $y$ due to the translational symmetry along this direction.

\subsubsection{Probability density and probability current for $f(1)\neq0$}
For the BGCS of Eq. (\ref{eq.coherente11}), with $f(n)=1$, we obtain the following probability density and probability current, respectively: 

\begin{equation}
\begin{split}
 \rho_{\alpha}(x,r,\theta) = \frac{1}{2e^{r^{2}} - r^{2} - 1} \left\lbrace  \displaystyle \sum_{n=2}^\infty \sum_{m=2}^\infty \frac{r^{n+m} \cos[(n-m)\theta] }{\sqrt{n! m!}} \, \rho_{n,m}(x) +  \left[ \psi_{0}(x)\right]  ^{2} + r^{2}\left[ \psi_{1}(x)\right] ^{2} \right. \\ 
\left. + 2r \cos\theta\, \psi_{0}(x) \psi_{1}(x) + 2 \displaystyle \sum_{n=2}^\infty \frac{r^{n}}{\sqrt{n!}} \left[  \cos(n\theta)\, \psi_{0}(x)+ r\cos\left[ (n-1)\theta\right] \, \psi_{1}(x)\right]  \psi_{n}(x) \right\rbrace , \label{eq.prob1}
\end{split}
\end{equation}

\begin{equation}
\begin{split}
J_{\alpha,x}(x,r,\theta)=&\frac{\hbar}{m^{*}(2e^{r^{2}}-r^{2}- 1)} \left\lbrace \sum_{n=2}^\infty \sum_{m=2}^\infty \frac{r^{n+m} \sin[(n-m)\theta]}{\sqrt{n!m!}}
\left\lbrace  \psi_{m}(x)\left[ \sqrt{(n-2)\omega} \, \psi_{n-3}(x) \right. \right. \right. \\  &
\left.\left. \left. -\left( \dfrac{\omega x}{2}+2k\right) \psi_{n-2}(x) \right]+\psi_{m-2}(x)\left[ \sqrt{n\omega} \, \psi_{n-1}(x)-\left( \dfrac{\omega x}{2}\right) \psi_{n}(x) \right]  \right\rbrace  \right. \\ &
\left. + \displaystyle \sum_{n=2}^\infty \frac{r^{n}}{\sqrt{n!}}
\left[ \sqrt{(n-2)\omega} \, \psi_{n-3}(x)-\left( \dfrac{\omega x}{2}+2k\right) \psi_{n-2}(x) \right]\left[\psi_{0}(x)\sin(n\theta) \displaystyle \right.\right.  \\ &
\left.\left. \displaystyle +\psi_{1}(x)r\sin\left[ (n-1)\theta\right]  \right]  - \sum_{m=2}^\infty \frac{r^{m}}{\sqrt{m!}}
\left\lbrace \sqrt{\omega} \, \psi_{0}(x)\psi_{m-2}(x)r \sin[(m-1)\theta] \right. \right. \\ & 
\left. \left. - \left(\dfrac{\omega x}{2} \right) \psi_{m-2}(x) \left[ \psi_{0}(x)\sin(m\theta)+\psi_{1}(x)r\sin\left[ (m-1)\theta\right]  \right]\right\rbrace  
\right\rbrace. \label{eq.corr1x}
\end{split}
\end{equation}

\begin{equation}
\begin{split}
J_{\alpha,y}(x,r,\theta)=&\frac{\hbar}{m^{*}(2e^{r^{2}}-r^{2}- 1)} \left\lbrace \sum_{n=2}^\infty \sum_{m=2}^\infty \frac{r^{n+m} \cos[(n-m)\theta]}{\sqrt{n!m!}}
\left\lbrace  \psi_{m}(x)\left[ \sqrt{(n-2)\omega} \, \psi_{n-3}(x) \right. \right. \right. \\  &
\left.\left. \left. -\left( \dfrac{\omega x}{2}+2k\right) \psi_{n-2}(x) \right]-\psi_{m-2}(x)\left[ \sqrt{n\omega} \, \psi_{n-1}(x)-\left( \dfrac{\omega x}{2}\right) \psi_{n}(x) \right]  \right\rbrace  \right. \\ &
\left. + \displaystyle \sum_{n=2}^\infty \frac{r^{n}}{\sqrt{n!}}
\left[ \sqrt{(n-2)\omega} \, \psi_{n-3}(x)-\left( \dfrac{\omega x}{2}+2k\right) \psi_{n-2}(x) \right]\left[\psi_{0}(x)\cos(n\theta) \displaystyle \right.\right.  \\ &
\left.\left. \displaystyle +\psi_{1}(x)r\cos\left[ (n-1)\theta\right]  \right]  - \sum_{m=2}^\infty \frac{r^{m}}{\sqrt{m!}}
\left\lbrace \sqrt{\omega} \, \psi_{0}(x)\psi_{m-2}(x)r \cos[(m-1)\theta] \right. \right. \\ & 
\left. \left. - \left(\dfrac{\omega x}{2} \right) \psi_{m-2}(x) \left[ \psi_{0}(x)\cos(m\theta)+\psi_{1}(x)r\cos\left[ (m-1)\theta\right]  \right]\right\rbrace  
\right\rbrace. \label{eq.corr1y}
\end{split}
\end{equation}

\begin{figure}[t]
\begin{center} 
  \includegraphics[width=15.5cm, height=11cm]{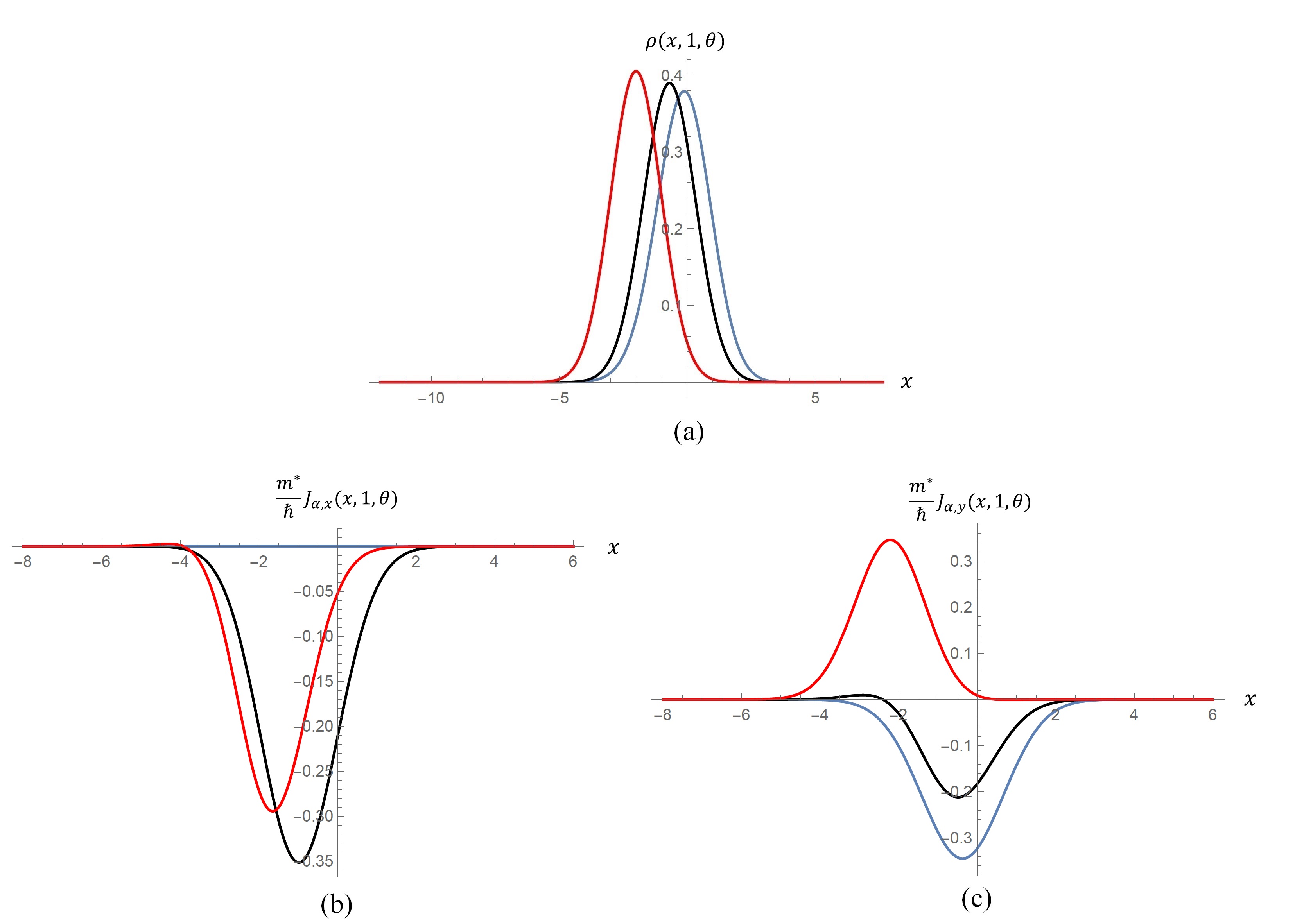}
   \caption{(a) Probability density $\rho_{\alpha}(x)$ and (b,c) probability currents $\left( m^{*} /\hslash\right) \,J_{\alpha, x, y}(x)$ for $f(n)=1$ with $\omega=1$, $k=1$ and $r=1$. The blue, black and red colors correspond to $\theta=\left\lbrace 0, \pi/4, \pi/2\right\rbrace$, respectively.}
  \label{fig.prob1}
  \end{center}  
\end{figure}

Some graphs of these quantities, for different values of $\theta$, are shown in Figure \ref{fig.prob1}.

\subsubsection{Probability density and probability current for $f(1)=0$}
\subsubsection*{A. Case with $f(2)\neq0$. }
For the case when $f(n)=\sqrt{n-1}/\sqrt{n}$ the BGCS of Eq. (\ref{eq.coherente22}) lead to
\begin{equation}
\begin{split}
\rho_{\alpha}(x,r,\theta)= \frac{1}{2e^{r^{2}} - 1}\left\lbrace \left[ \psi_{1}(x)\right]^{2} + \sum_{n=1}^\infty \sum_{m=1}^\infty \frac{r^{n+m} \cos[(n-m)\theta]}{\sqrt{n!m!}}\, \rho_{n+1,m+1}(x) \right. \\ 
\left.+ 2\sum_{n=1}^\infty \frac{r^{n}\cos(n\theta)}{\sqrt{n!}} \psi_{1}(x) \psi_{n+1}(x)\right\rbrace, \label{eq.prob2}
\end{split}
\end{equation}

\begin{equation}
\begin{split}
J_{\alpha,x}(x,r,\theta)&= \dfrac{\hbar}{m^{*}(2e^{r^{2}} - 1)} \left\lbrace \displaystyle \sum_{n=1}^\infty \sum_{m=1}^\infty \frac{r^{n+m} \sin[(n-m)\theta]}{\sqrt{n!m!}}
\left\lbrace  \psi_{m+1}(x)\left[ \sqrt{(n-1)\omega} \, \psi_{n-2}(x) \right. \right. \right. \\  &
\left.\left. \left. - \left( \dfrac{\omega x}{2}+2k\right) \psi_{n-1}(x) \right]+\psi_{m-1}(x)\left[ \sqrt{(n+1)\omega} \, \psi_{n}(x)-\left( \dfrac{\omega x}{2}\right) \psi_{n+1}(x) \right]  \right\rbrace  \right. \\ &
\left. + \displaystyle \sum_{n=1}^\infty \frac{r^{n} \sin(n\theta)}{\sqrt{n!}} \psi_{1}(x) \left[ \sqrt{(n-1)\omega} \, \psi_{n-2}(x)-\left( \dfrac{\omega x}{2}+2k\right) \psi_{n-1}(x) \right]  
 \right. \\ &
\left. - \displaystyle \sum_{m=1}^\infty \frac{r^{m} \sin(m\theta)}{\sqrt{m!}}
 \psi_{m-1}(x) \left[ \sqrt{\omega} \, \psi_{0}(x)-\left( \dfrac{\omega x}{2}\right) \psi_{1}(x) \right]  
\right\rbrace, \label{eq.corr2x}
\end{split}
\end{equation}

\begin{equation}
\begin{split}
J_{\alpha,y}(x,r,\theta)&= \dfrac{\hbar}{m^{*}(2e^{r^{2}} - 1)} \left\lbrace \displaystyle \sum_{n=1}^\infty \sum_{m=1}^\infty \frac{r^{n+m} \cos[(n-m)\theta]}{\sqrt{n!m!}} \left\lbrace  \psi_{m+1}(x)\left[ \sqrt{(n-1)\omega} \, \psi_{n-2}(x) \right. \right. \right. \\  &
\left.\left. \left. - \left( \dfrac{\omega x}{2}+2k\right) \psi_{n-1}(x) \right]-\psi_{m-1}(x)\left[ \sqrt{(n+1)\omega} \, \psi_{n}(x)-\left( \dfrac{\omega x}{2}\right) \psi_{n+1}(x) \right]  \right\rbrace  \right. \\ &
\left. + \displaystyle \sum_{n=1}^\infty \frac{r^{n} \cos(n\theta)}{\sqrt{n!}}
  \psi_{1}(x) \left[ \sqrt{(n-1)\omega} \, \psi_{n-2}(x)-\left( \dfrac{\omega x}{2}+2k\right) \psi_{n-1}(x) \right]  
 \right. \\ &
\left. - \displaystyle \sum_{m=1}^\infty \frac{r^{m} \cos(m\theta)}{\sqrt{m!}}
 \psi_{m-1}(x) \left[ \sqrt{\omega} \, \psi_{0}(x)-\left( \dfrac{\omega x}{2}\right) \psi_{1}(x) \right]  
\right\rbrace. \label{eq.corr2y}
\end{split}
\end{equation}

\begin{figure}[t]
\begin{center} 
  \includegraphics[width=16.2cm, height=11cm]{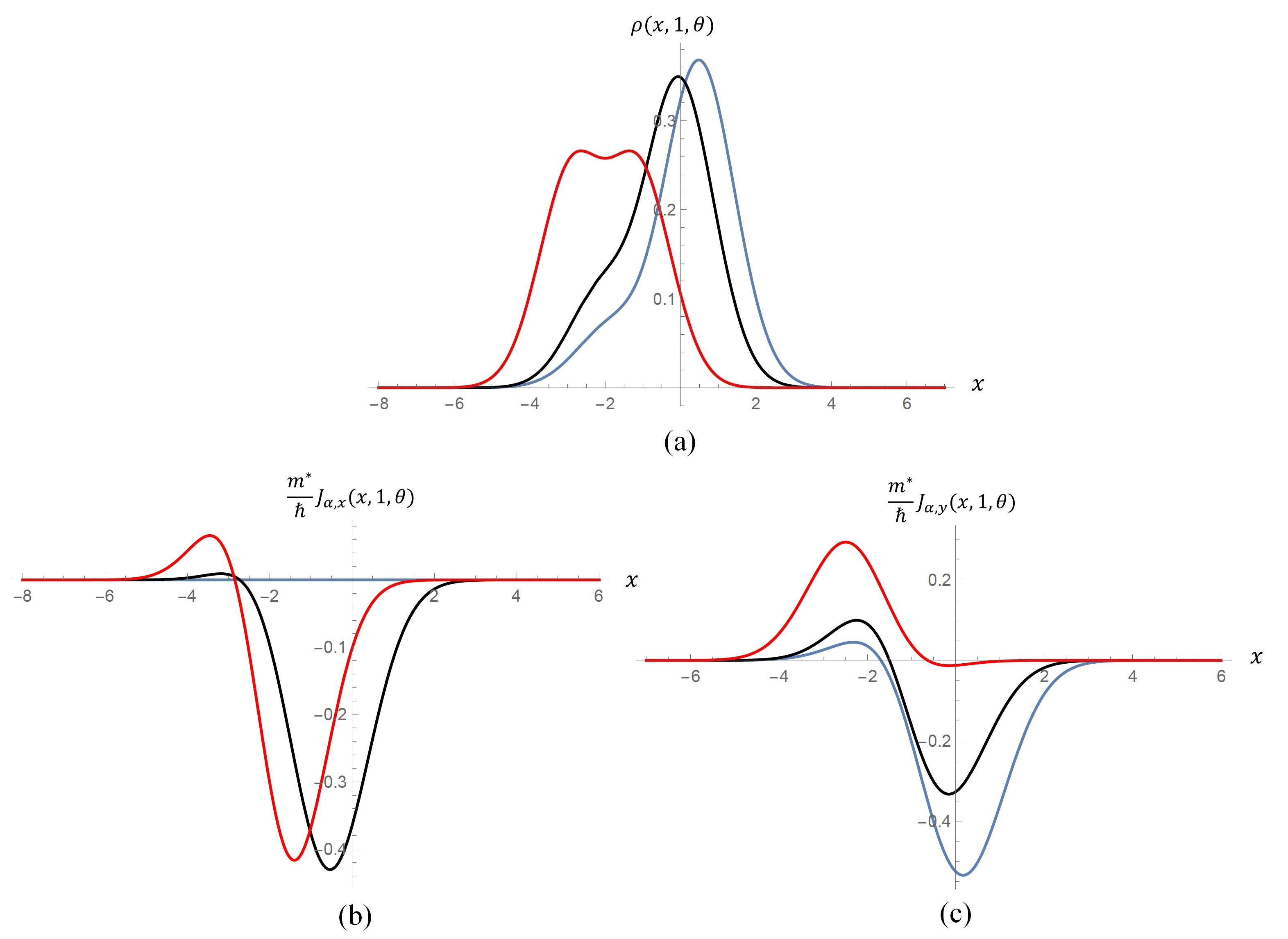}
   \caption{(a) Probability density $\rho_{\alpha}(x)$ and (b,c) probability currents $\left( m^{*} /\hslash\right) \,J_{\alpha, x, y}(x)$ for $f(n)=\sqrt{n-1}/\sqrt{n}$ with $\omega=1$, $k=1$ and $r=1$. The blue, black and red colors correspond to $\theta=\left\lbrace 0, \pi/4, \pi/2\right\rbrace$, respectively.} 
  \label{fig.prob2}
  \end{center}  
\end{figure}

Some graphs of these quantities, for different values of $\theta$, are shown in Figure \ref{fig.prob2}.

\subsubsection*{B. Case with $f(2)=0$.}

Finally, when $f(n)=(n-2)\sqrt{n-1}/\sqrt{n}$ the BGCS of Eq. (\ref{eq.coherente33}) produce the following probability density and probability current

\begin{equation}
\rho_{\alpha}(x,r,\theta) =  \frac{1}{2\,{_0}F_{2}(1,2;r^{2})}  \sum_{n=0}^\infty \sum_{m=0}^\infty \frac{r^{n+m} \cos[(n-m)\theta]}{n! \, m! \sqrt{(n+1)! \, (m+1)!}}\, \rho_{n+2,m+2}(x), \label{eq.prob3}
\end{equation}

\begin{figure}[t]
\begin{center} 
  \includegraphics[width=15.5cm, height=11cm]{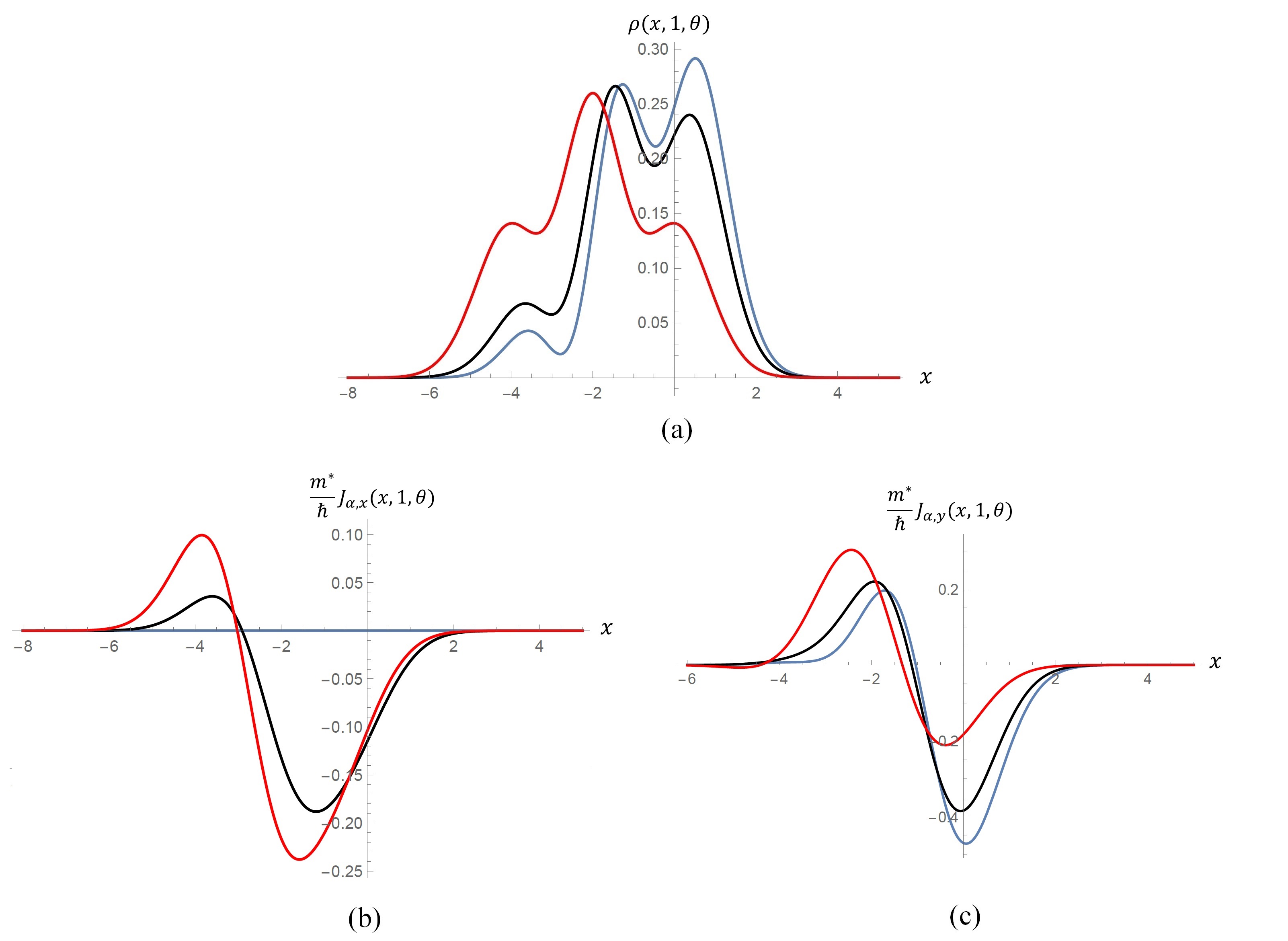}
   \caption{(a) Probability density $\rho_{\alpha}(x)$ and (b,c) probability currents $\left( m^{*} /\hslash\right) \,J_{\alpha, x, y}(x)$ for $f(n)=(n-2)\sqrt{n-1}/\sqrt{n}$ with $\omega=1$, $k=1$ and $r=1$. The blue, black and red colors correspond to $\theta=\left\lbrace 0, \pi/4, \pi/2\right\rbrace$, respectively.}
  \label{fig.prob3}
  \end{center}  
\end{figure}

\begin{equation}
\begin{split}
J_{\alpha,x}(x,r,\theta)&= \frac{\hbar}{2m^{*} {_0}F_{2}(1,2;r^{2})} \displaystyle \sum_{n=0}^\infty \sum_{m=0}^\infty \frac{r^{n+m} \sin[(n-m)\theta]}{n! \, m! \sqrt{(n+1)! \, (m+1)!}} \left\lbrace  \psi_{m+2}(x)\left[ \sqrt{n\omega} \, \psi_{n-1}(x)\right.\right.  \\ &
\left. \left. \,\ \,\ \,\ \,\  -\left( \dfrac{\omega x}{2}+2k\right)\psi_{n}(x) \right] +\psi_{m}(x)\left[ \sqrt{(n+2)\omega} \, \psi_{n+1}(x)-\left( \dfrac{\omega x}{2}\right) \psi_{n+2}(x) \right] \right\rbrace, \label{eq.corr3x}
\end{split}
\end{equation}

\begin{equation}
\begin{split}
J_{\alpha,y}(x,r,\theta)&= \frac{\hbar}{2m^{*} {_0}F_{2}(1,2;r^{2}) }  \displaystyle \sum_{n=0}^\infty \sum_{m=0}^\infty \frac{r^{n+m} \cos[(n-m)\theta]}{n! \, m! \sqrt{(n+1)! \, (m+1)!}} \left\lbrace  \psi_{m+2}(x)\left[ \sqrt{n\omega} \, \psi_{n-1}(x)\right.\right.  \\ &
\left. \left.  \,\ \,\ \,\ \,\ -\left( \dfrac{\omega x}{2}+2k\right) \psi_{n}(x) \right] -\psi_{m}(x)\left[ \sqrt{(n+2)\omega} \, \psi_{n+1}(x)-\left( \dfrac{\omega x}{2}\right) \psi_{n+2}(x) \right] \right\rbrace. \label{eq.corr3y}
\end{split}
\end{equation}

Some graphs of these quantities, for different values of $\theta$, are shown in Figure \ref{fig.prob3}.
\\
\\
\\
\\

As can be seen in Figures (\ref{fig.prob1}\,-\,\ref{fig.prob3}) the maximum of the probability density moves along $x$ direction when $\theta$ increases. On the other hand, non-null probability currents along both  $x$ and $y$ directions are obtained for our BGCS. However, for $\theta=0$ only current along $y$\,-\,direction appears, as it happens for the monolayer graphene coherent states.

\subsection{Mean energy value}
The energy of the system is another quantity useful to characterize the coherent states. In addition, in the next section we will use it to explain the time evolution of the monolayer graphene coherent states, as well as of our bilayer graphene coherent states. This quantity is calculated as follows:
\begin{equation}
E= \langle \hat{H}\rangle_{\alpha}=\langle \Psi_{\alpha} \vert \hat{H} \vert \Psi_{\alpha} \rangle, \label{eq.espenergia}
\end{equation}
i.e., it is the mean value of the effective Hamiltonian of Eq. (\ref{eq.ham4}) in the states $\Psi_{\alpha}(x,y)$. 
We will calculate next this mean value for each set of coherent states previously derived.

\begin{figure}[t]
\begin{center} 
  \includegraphics[width=15cm]{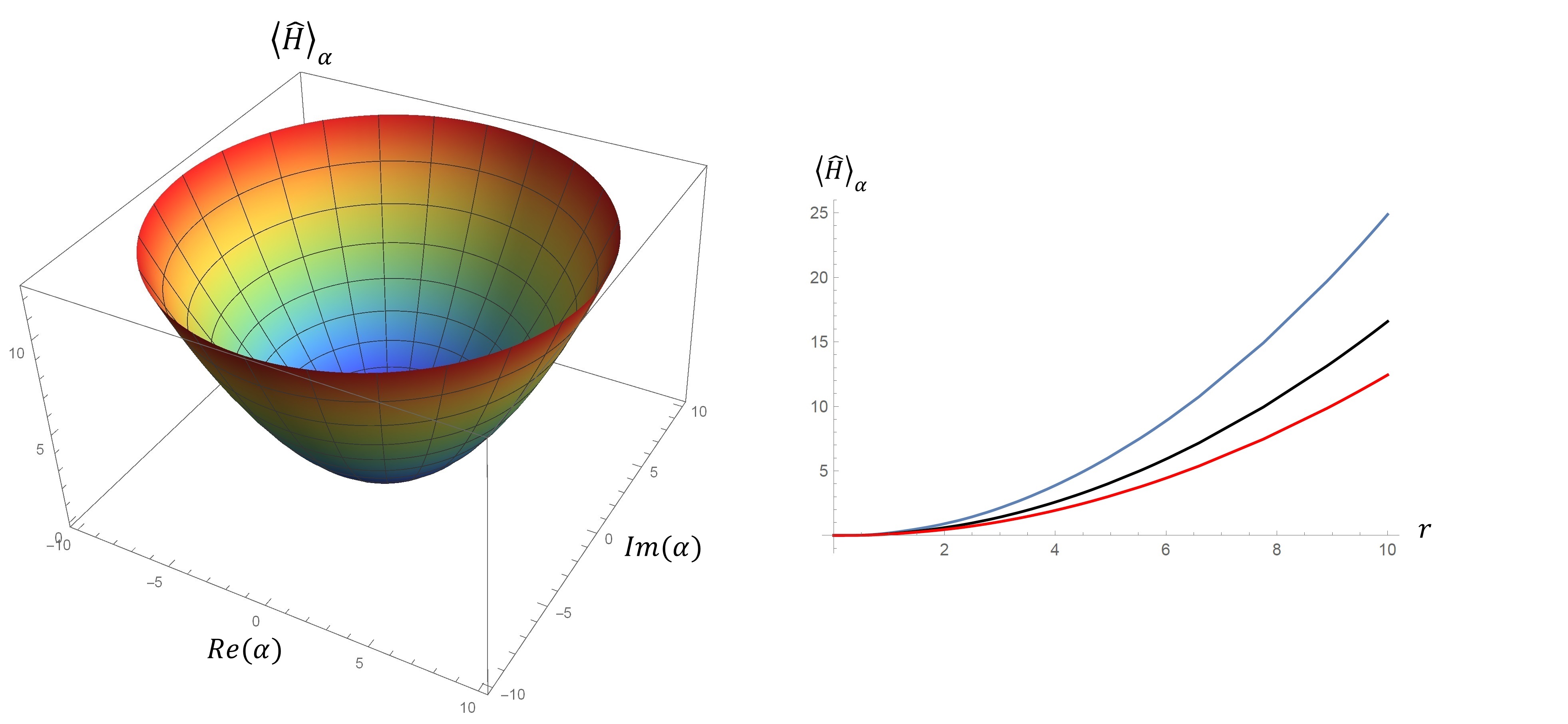}
   \caption{Left: Mean energy as function of $\alpha$ for the BGCS with $f(n)=1$ and $B=1/8$. Right: Mean energy as function of $r=\vert \alpha\vert$ for different magnetic field intensities: the blue, black and red colors correspond to $B=\left\lbrace 1/4, \, 1/6, \,  1/8\right\rbrace$, respectively.}  
  \label{fig.ener1}
  \end{center}  
\end{figure}

\subsubsection{ $\langle \hat{H}\rangle_{\alpha}$ for $f(1)\neq0$}
For the coherent states of Eq. (\ref{eq.coherente11}) we obtain (see Figure \ref{fig.ener1})
\begin{equation}
\langle \Psi_{\alpha} \vert \hat{H} \vert \Psi_{\alpha} \rangle = \left( \frac{2\hbar \omega^{*}_{c}}{2e^{r^{2}} - r^{2} - 1}\right) \sum_{n=2}^\infty  \frac{\sqrt{n(n-1)} \, r^{2n}}{n!}. \label{eq.ener1}
\end{equation}

\subsubsection{$\langle \hat{H}\rangle_{\alpha}$ for $f(1)=0$}
\subsubsection*{A. Case with $f(2)\neq0$.}
\begin{figure}[t]
\begin{center} 
  \includegraphics[width=15cm, height=7.6cm]{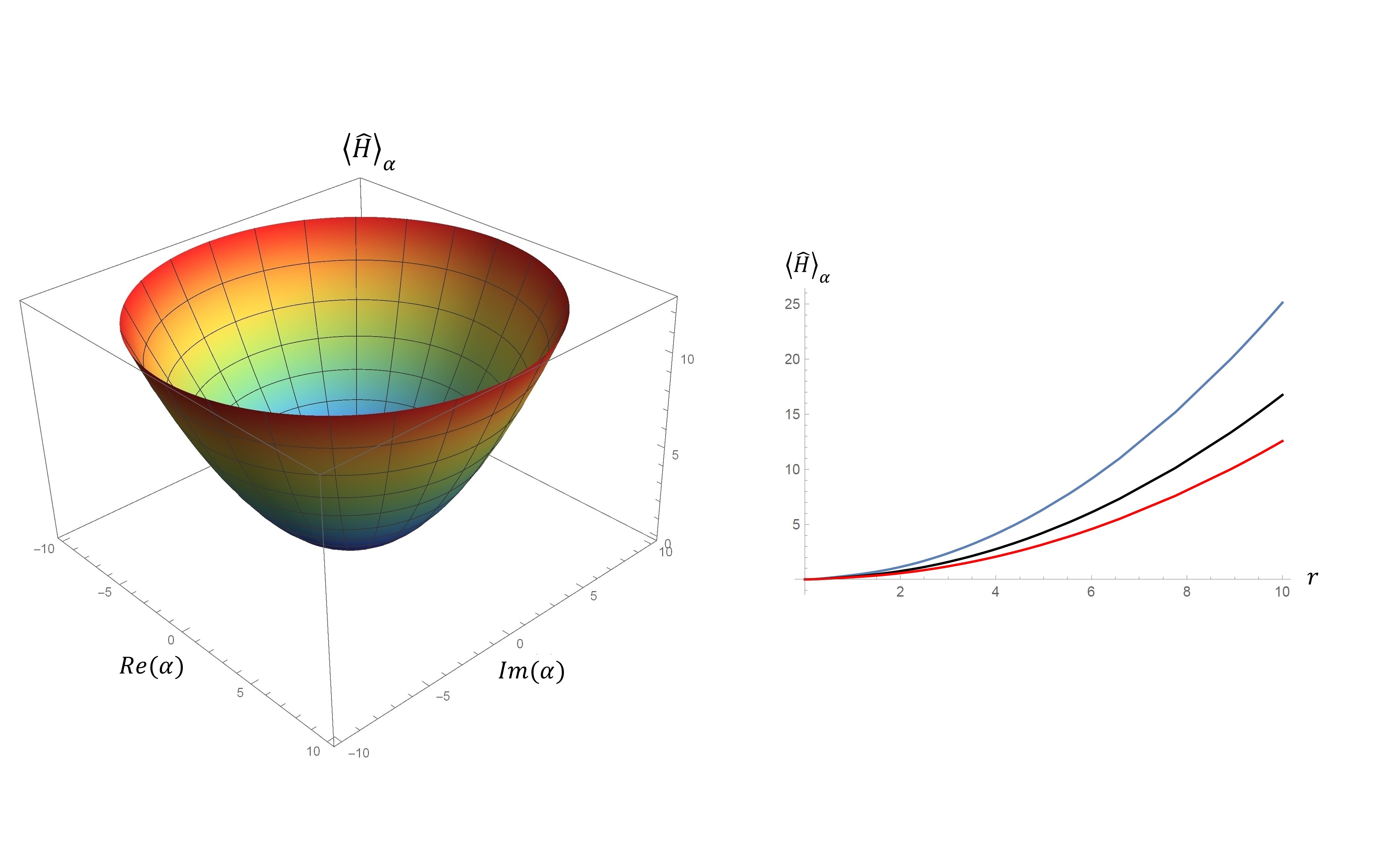}
   \caption{Left: Mean energy as function of $\alpha$ for the BGCS with $f(n)=\sqrt{n-1}/\sqrt{n}$ and $B=1/8$. Right: Mean energy as function of $r=\vert \alpha\vert$ for different magnetic field intensities: the blue, black and red colors correspond to $B=\left\lbrace 1/4, \, 1/6, \,  1/8\right\rbrace $, respectively.}  
  \label{fig.ener2}
  \end{center}  
\end{figure}

The coherent states of Eq. (\ref{eq.coherente22}) lead to (see Figure \ref{fig.ener2})

\begin{equation}
\langle \Psi_{\alpha} \vert \hat{H} \vert \Psi_{\alpha} \rangle = \left( \frac{2\hbar \omega^{*}_{c}}{2e^{r^{2}} - 1}\right) \sum_{n=1}^\infty  \frac{\sqrt{n(n+1)} \, r^{2n}}{n!}. \label{eq.ener2}
\end{equation}

\subsubsection*{B. Case with $f(2)=0$.}
\begin{figure}[!]
\begin{center} 
  \includegraphics[width=15cm, height=7.4cm]{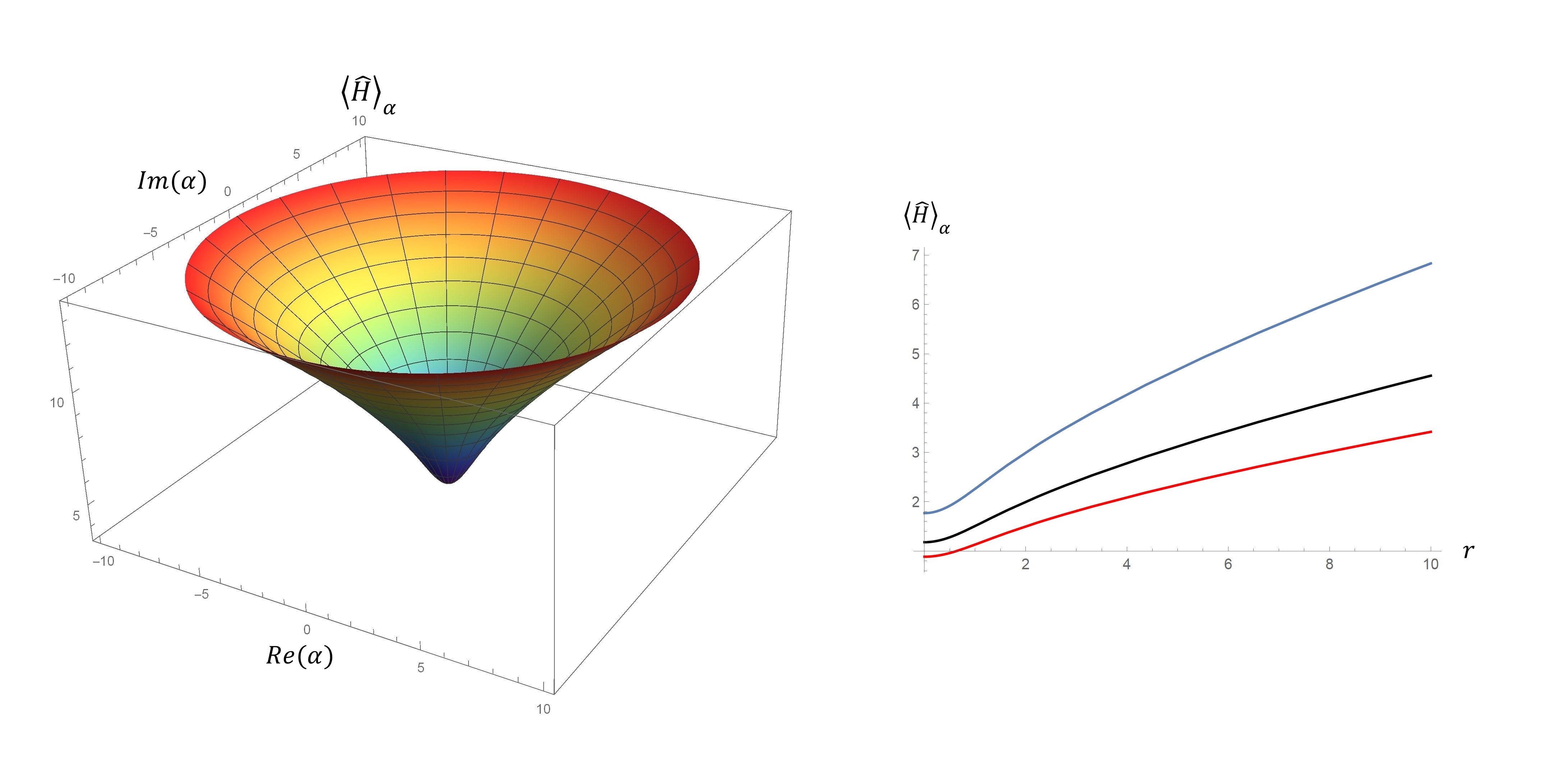}
   \caption{Left: Mean energy as function of $\alpha$ for the BGCS with $f(n)=(n-2)\sqrt{n-1}/\sqrt{n}$  and $B=1/8$. Right:  Mean energy as function of $r=\vert \alpha\vert$ for different magnetic field intensities: the blue, black and red colors correspond to $B=\left\lbrace 1/4, \, 1/6, \,  1/8\right\rbrace $, respectively.} 
  \label{fig.ener3}
  \end{center}  
\end{figure}

Finally, for the coherent states of Eq. (\ref{eq.coherente33}) we arrive to (see Figure \ref{fig.ener3})

\begin{equation}
\langle \Psi_{\alpha} \vert \hat{H} \vert \Psi_{\alpha} \rangle = \left( \frac{\hbar \omega^{*}_{c}}{\,\ {_0}F_{2}(1,2;r^{2})} \right) \sum_{n=0}^\infty  \frac{\sqrt{(n+1)(n+2)} \, r^{2n}}{(n!)^{2} \, (n+1)!}. \label{eq.ener3}
\end{equation}
\\
\\
\\
\\
\\

In Figures (\ref{fig.ener1}\,-\,\ref{fig.ener3}) we are showing the mean energy value for the three sets of BGCS built previously. As can be seen, it is a growing function of $\alpha$, but in the last case (Eq. (\ref{eq.ener3})) it grows more slowly than in the other two cases, since the structure is different for each set of coherent states. Finally, we can also note that $\langle \hat{H}\rangle_{\alpha}$ grows as the magnetic field intensity does, and this increase is proportional to $B$.

\section{Time evolution of the graphene coherent states}
The time evolution of a quantum state is obtained by acting the unitary operator $\hat{U}(t,t_{0})$ as follows \cite{zettili}
\begin{equation}
\vert \Psi(t) \rangle=\hat{U}(t,t_{0})\vert \Psi(t_{0}) \rangle, \qquad t>t_{0}, \label{eq.uniope}
\end{equation}
where $\hat{U}(t,t_{0})$ is known as the \textit{evolution operator}. Note that
\begin{equation}
\hat{U}(t_{0},t_{0})=\hat{I}, \label{eq.uniope1}
\end{equation}
with $\hat{I}$ being the identity operator.
\\

If we substitute Eq. (\ref{eq.uniope}) into the time-dependent Schr\"odinger equation we get

\begin{equation}
i\hbar\dfrac{\partial \hat{U}(t,t_{0})}{\partial t}=\hat{H}\hat{U}(t,t_{0}). \label{eq.timedepend}
\end{equation}
In particular, for time-independent Hamiltonians the last equation can be simply integrated, with the initial condition of Eq. (\ref{eq.uniope1}), in order to obtain:

\begin{equation}
\hat{U}(t_{0},t_{0})=\exp\left[ -i(t-t_{0})\hat{H}/\hbar \right]\qquad \Rightarrow \qquad \vert \Psi(t) \rangle=\exp\left[-i(t-t_{0})\hat{H}/\hbar \right]\vert \Psi(t_{0}) \rangle. \label{eq.evolucion}
\end{equation}

\subsection{Evolution of the bilayer graphene coherent states}

\subsubsection{Evolution of the BGCS for $f\left( 1\right) \neq 0$}
One of the most important properties of the standard coherent states is their stability under time-evolution i.e., a standard coherent state evolves into a standard coherent state at any time \cite{Gerry}. We calculate next the time evolution of the BGCS of Eq. (\ref{eq.coherente11}) for $t_{0}=0$ (see Eq. (\ref{eq.evolucion})), which leads to
\begin{equation}
\Psi_{\alpha}(x,y;t)= \frac{1}{\sqrt{2e^{r^{2}} - r^{2} - 1}} \left[  \Psi_{0}
(x,y)+\alpha\Psi_{1}(x,y)+ \sum_{n=2}^\infty  \frac{\sqrt{2} \alpha^{n}}{\sqrt{n!}} e^{ -i\omega^{*}_{c}\sqrt{n(n-1)} t} \, \Psi_{n}(x,y) \right]. \label{eq.evolbic1}
\end{equation}

\subsubsection{Evolution of the BGCS for $f\left( 1\right)=0$}
\subsubsection*{A. Case with $f\left( 2\right) \neq 0$.}
For the states of Eq. (\ref{eq.coherente22}) it is obtained  
\begin{equation}
\Psi_{\alpha}(x,y;t)= \frac{1}{\sqrt{2e^{r^{2}}- 1}} \left[\Psi_{1}(x,y)+ \sum_{n=2}^\infty  \frac{\sqrt{2}\alpha^{n-1}}{\sqrt{(n-1)!}} e^{-i\omega^{*}_{c}\sqrt{n(n-1)}t } \, \Psi_{n}(x,y) \right]. \label{eq.evolbic2}
\end{equation}

\subsubsection*{B. Case with $f\left( 2\right)=0$.}
Finally, for the BGCS of Eq. (\ref{eq.coherente33}) we arrive at
\begin{equation}
\Psi_{\alpha}(x,y;t)= \frac{1}{\sqrt{{_0}F_{2}(1,2;r^{2})}} \left[ \sum_{n=2}^\infty  \frac{\alpha^{n-2}}{(n-2)!\sqrt{(n-1)!}} e^{-i\omega^{*}_{c}\sqrt{n(n-1)}t } \, \Psi_{n}(x,y)\right]. \label{eq.evolbic3}  
\end{equation}

\begin{figure}[t]
\begin{center} 
  \includegraphics[width=16cm]{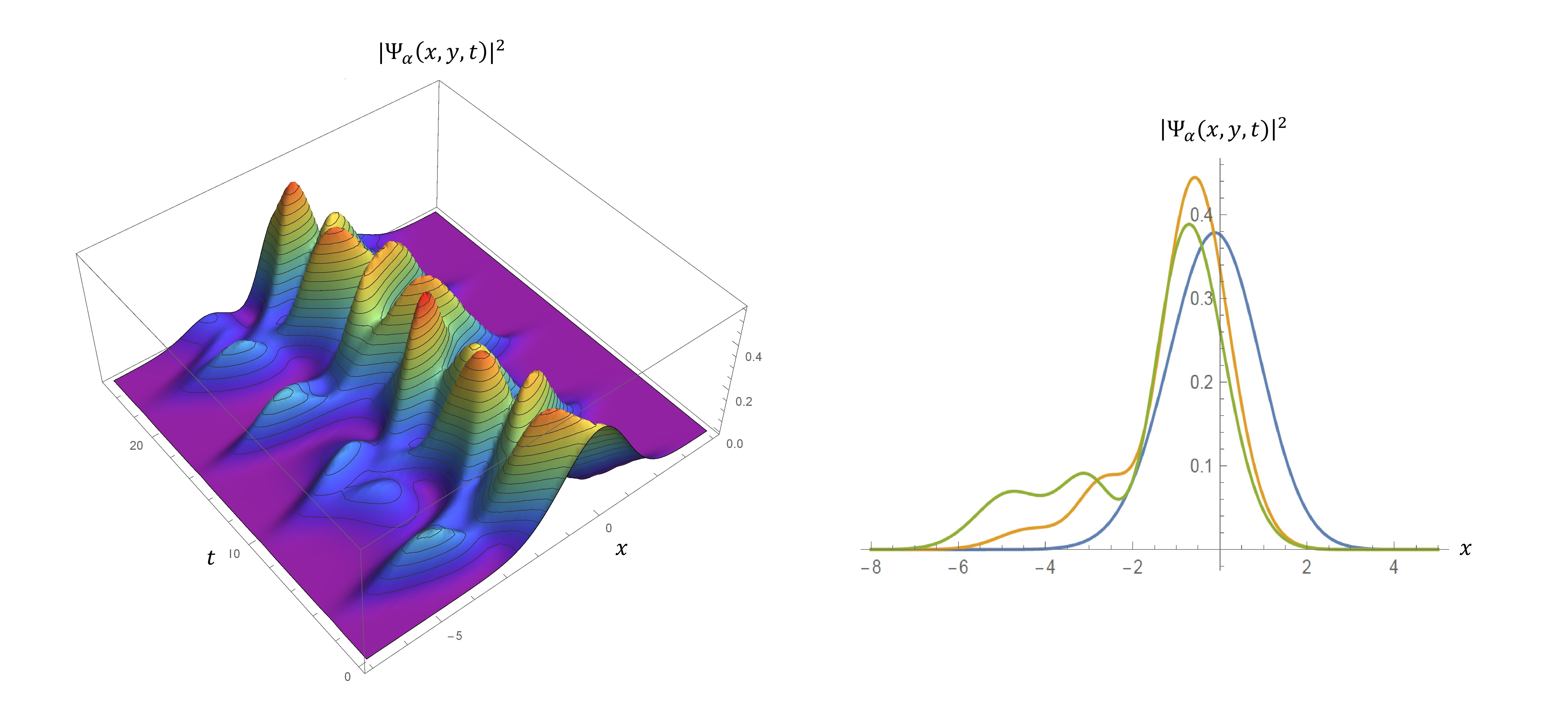}
   \caption{Left: Probability density $|\Psi_{\alpha}(x,y;t)|^{2}$ for the BGCS with $f(n)=1$ (Eq. (\ref{eq.evolbic1})), $r=1$, $\theta=0$ and $\omega_{c}^{*}=1$. Right: Probability density $|\Psi_{\alpha}(x,y;t)|^{2}$ for some fixed times (the suggested approximate period and some of its multiples). The blue, green and orange lines correspond to $\tau=\left\lbrace 0, \sqrt{2}\pi, 2\sqrt{2}\pi\right\rbrace $, respectively.} 
  \label{fig.denbica1}
  \end{center}  
\end{figure}

\begin{figure}
\begin{center}
  \includegraphics[width=16cm]{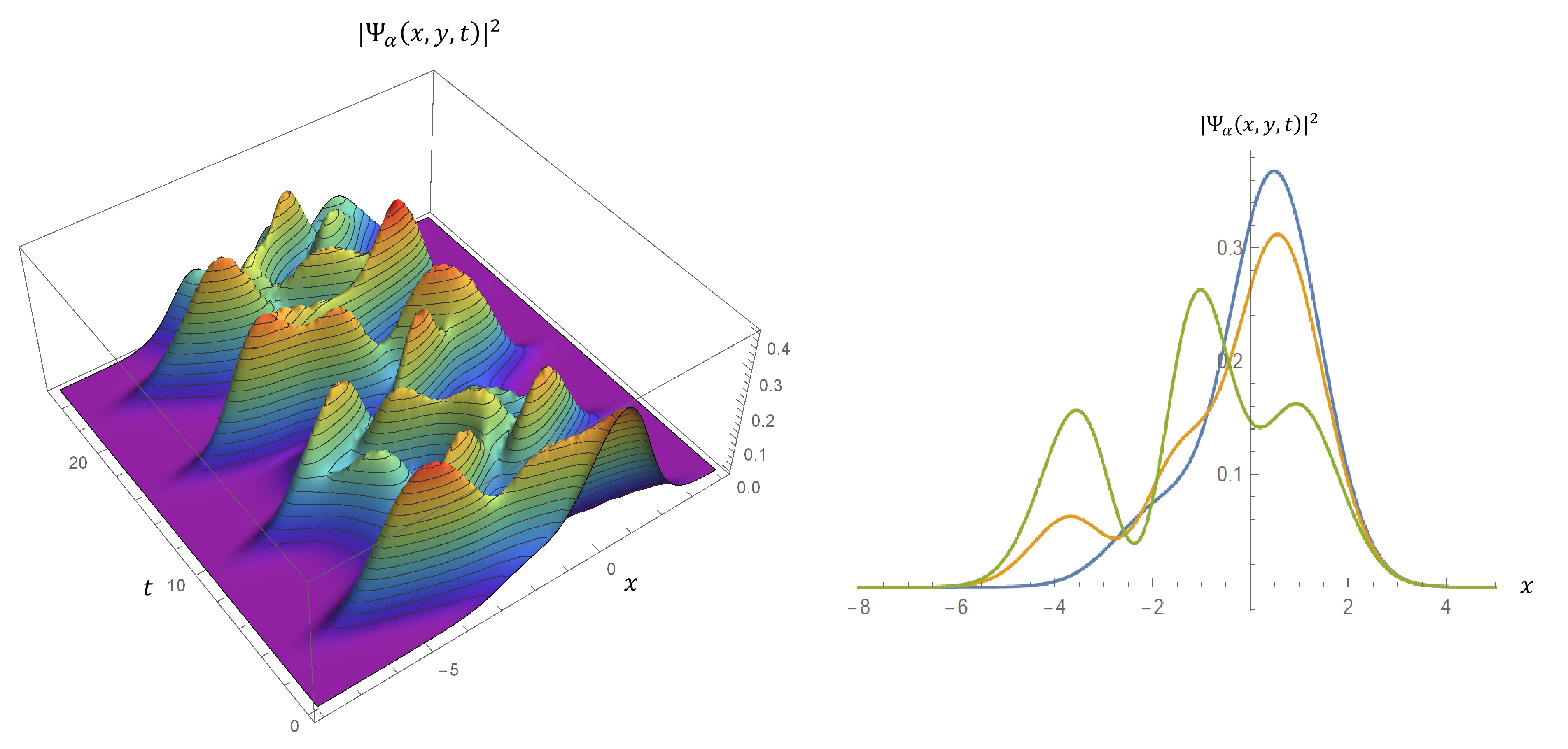}
   \caption{Left: Probability density  $|\Psi_{\alpha}(x,y;t)|^{2}$ for the BGCS with $f(n)=\sqrt{n-1}/\sqrt{n}$ (Eq. (\ref{eq.evolbic2})), $r=1$, $\theta=0$ and $\omega_{c}^{*}=1$. Right: Probability density $|\Psi_{\alpha}(x,y;t)|^{2}$  for some fixed times (the suggested approximate period and some of its multiples). The blue, green and orange lines correspond to $\tau= \left\lbrace 0, 2\pi, 4 \pi\right\rbrace $, respectively.}   
  \label{fig.denbica2}
  \end{center}  
\end{figure}

\begin{figure}
\begin{center} 
  \includegraphics[width=16cm]{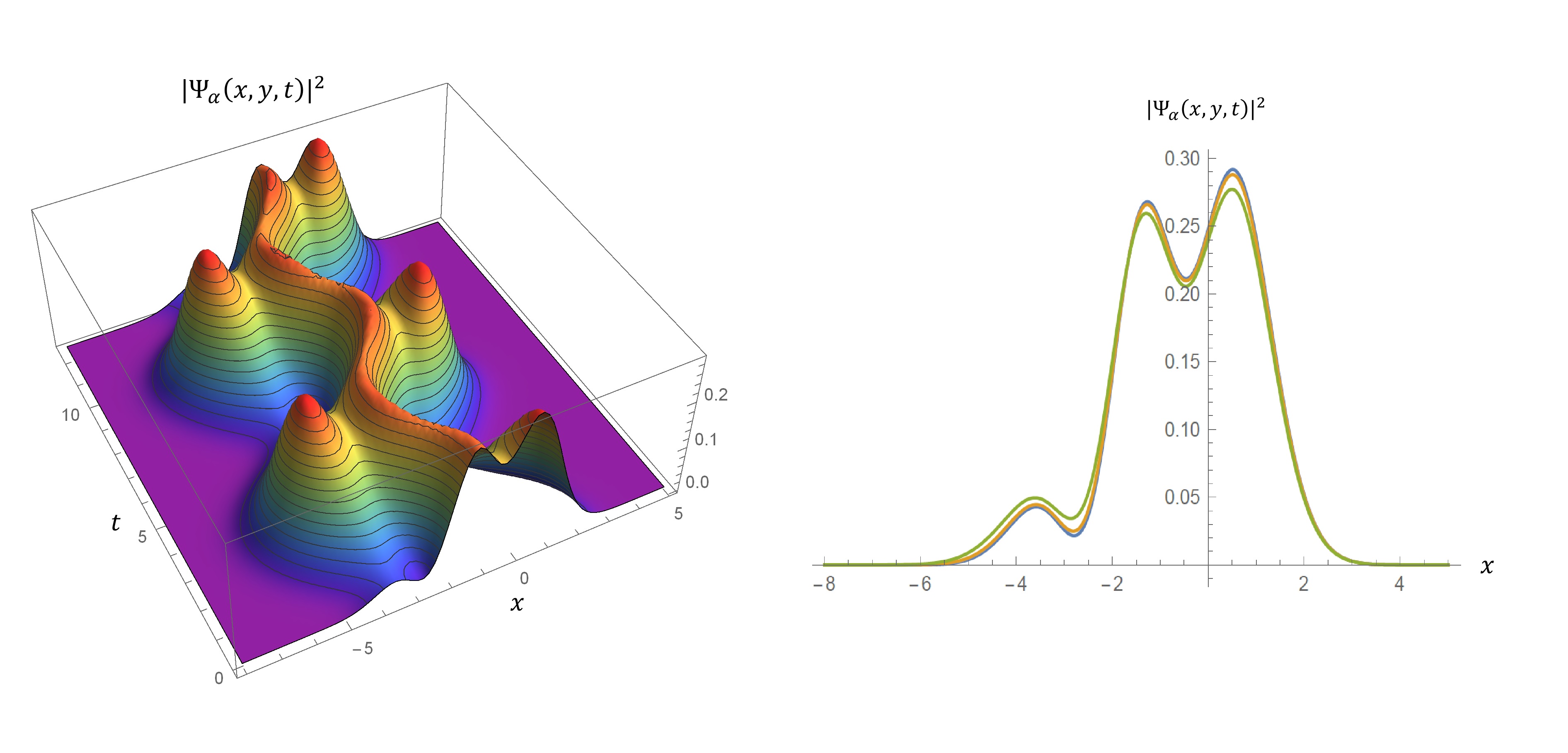}
 \caption{Left: Probability density $|\Psi_{\alpha}(x,y;t)|^{2}$ for the BGCS with $f(n)=(n-2)\sqrt{n-1}/\sqrt{n}$ (Eq. (\ref{eq.evolbic3})), $r=1$, $\theta=0$ and $\omega_{c}^{*}=1$. Right: Probability density $|\Psi_{\alpha}(x,y;t)|^{2}$ for some fixed times (the suggested approximate period and some of its multiples). The blue, green and orange lines correspond to $\tau=\left\lbrace 0, 2\pi, 4 \pi\right\rbrace $, respectively.}
  \label{fig.denbica3}
  \end{center}  
\end{figure}

The probability densities for the evolving states of Eqs. (\ref{eq.evolbic1}, \ref{eq.evolbic2}, \ref{eq.evolbic3}) are shown in  Figures \ref{fig.denbica1}, \ref{fig.denbica2} and \ref{fig.denbica3}, respectively.

\subsubsection{Discussion}
It is well known that the time stability of the standard coherent states comes from the fact that the energy levels for the harmonic oscillator are equally spaced \cite{Gerry}. For quantum systems without equally spaced energy levels this stability, in general, does not exist \cite{equidistant1, equidistant2,wu}. 
\\

For our BGCS the energy spectrum is not equidistant for all $n$ (see Eq. (\ref{eq.eigenvalues})). However, starting from a certain integer (for $n\gtrsim2$) this spectrum is practically lineal, thus making stable in time (to a good approximation) the BGCS for which the contribution of the eigenstates $\Psi_{0}(x,y)$ and $\Psi_{1}(x,y)$ is small compared with the contribution of all other eigenstates. This can be seen clearly in Figure \ref{fig.denbica3}, where the BGCS of Eq. (\ref{eq.evolbic3}) are stable in time, with the same period as the auxiliary harmonic oscillator ($\tau\simeq2\pi/\omega_{c}^{*}$). For the other two examples, Eqs. (\ref{eq.evolbic1}, \ref{eq.evolbic2}), the coherent states could involve the eigenstate $\Psi_{0}(x,y)$, $\Psi_{1}(x,y)$ or both in a non-trivial way, thus making that their time evolution in general would not be stable (see Figures \ref{fig.denbica1} and \ref{fig.denbica2}).
\\

Despite the BGCS of Eqs. (\ref{eq.evolbic1}, \ref{eq.evolbic2}) do not have always the period $2\pi/\omega_{c}^{*}$, however we propose a way to find a possible approximate period $\tau$ for these states. First of all let us note that, for a given $\alpha$, $\tau$ is closely related to the mean energy value and to the eigenvalues bounding this average. Hence, by setting $\alpha$ we must calculate first the value $\langle \hat{H}\rangle_{\alpha}$, then we must determine the interval in which it lies, bounded by two consecutive proper energies $E_{j+1}$ and $E_{j}$, i.e., $E_{j}<\langle \hat{H}\rangle_{\alpha}<E_{j+1}$. Finally, let us propose the next expression for the possible approximate period for the time-evolution of our BGCS:

\begin{equation}
\tau=\dfrac{2\pi\hbar}{E_{j+1}-E_{j}}. \label{eq.periodo}
\end{equation}

For the states of Eq. (\ref{eq.evolbic1}) with $\vert \alpha \vert=1$ (such that $0<\langle \hat{H}\rangle_{\alpha}= 0.76 \,\hbar \omega_{c}^{*}< E_{2}$) we obtain a possible approximate period $\tau\simeq\sqrt{2}\pi /\omega_{c}^{*}$ and for the states of Eq. (\ref{eq.evolbic2}) with $E_{2}<\langle \hat{H}\rangle_{\alpha}= 1.56 \,\hbar \omega_{c}^{*}< E_{3}$  we obtain $\tau\simeq2\pi / \omega_{c}^{*}$. In Figures \ref{fig.denbica1} and \ref{fig.denbica2} we also plot (right) the probability density for such a $\tau$ and some of its multiples in each case.

\subsection{Evolution of the monolayer graphene coherent states}
The monolayer graphene coherent states (MGCS) for a constant homogeneous magnetic field were recently derived in \cite{erick}. Let us calculate next the time evolution of these states, in similar cases that for bilayer graphene.
 
\subsubsection{Evolution of the MGCS for $f\left( 1\right) \neq 0$}
When taking $f(n)=1$ the evolving MGCS become

\begin{equation}
\Psi_{\alpha}(x,y;t)= \frac{1}{\sqrt{2e^{r^{2}}- 1}} \left[  \Psi_{0}
(x,y)+ \sum_{n=1}^\infty  \frac{\sqrt{2}\alpha^{n}}{\sqrt{n!}}  e^{-i v_{F}\sqrt{n \omega} \, t} \, \Psi_{n}(x,y) \right], \label{eq.evolmon1}
\end{equation}
where $\omega$ have dimensions of (lenght)$^{-2}$.

\subsubsection{Evolution of the MGCS for $f\left( 1\right)=0$}

\subsubsection*{A. Case with $f\left( 2\right) \neq 0$.}
For $f(n)=\sqrt{n-1}/\sqrt{n}$ it is obtained
\begin{equation}
\Psi_{\alpha}(x,y;t)= e^{-r^{2}/2} \sum_{n=1}^\infty  \frac{\alpha^{n-1}}{\sqrt{(n-1)!}}  e^{-i v_{F}\sqrt{n \omega} \, t} \, \Psi_{n}(x,y). \label{eq.evolmon2}   
\end{equation}

\subsubsection*{B. Case with $f\left( 2\right)=0$.}
For  $f(n)=(n-2)\sqrt{n-1}/\sqrt{n}$ we arrive at
\begin{equation}
\Psi_{\alpha}(x,y;t)= \frac{1}{\sqrt{{_0}F_{2}(1,2;r^{2})}} \sum_{n=2}^\infty  \frac{\alpha^{n-2}}{(n-2)!\sqrt{(n-1)!}} e^{-i v_{F}\sqrt{n \omega} \, t} \, \Psi_{n}(x,y). \label{eq.evolmon3}
\end{equation}

\begin{figure}
\begin{center} 
  \includegraphics[width=16cm]{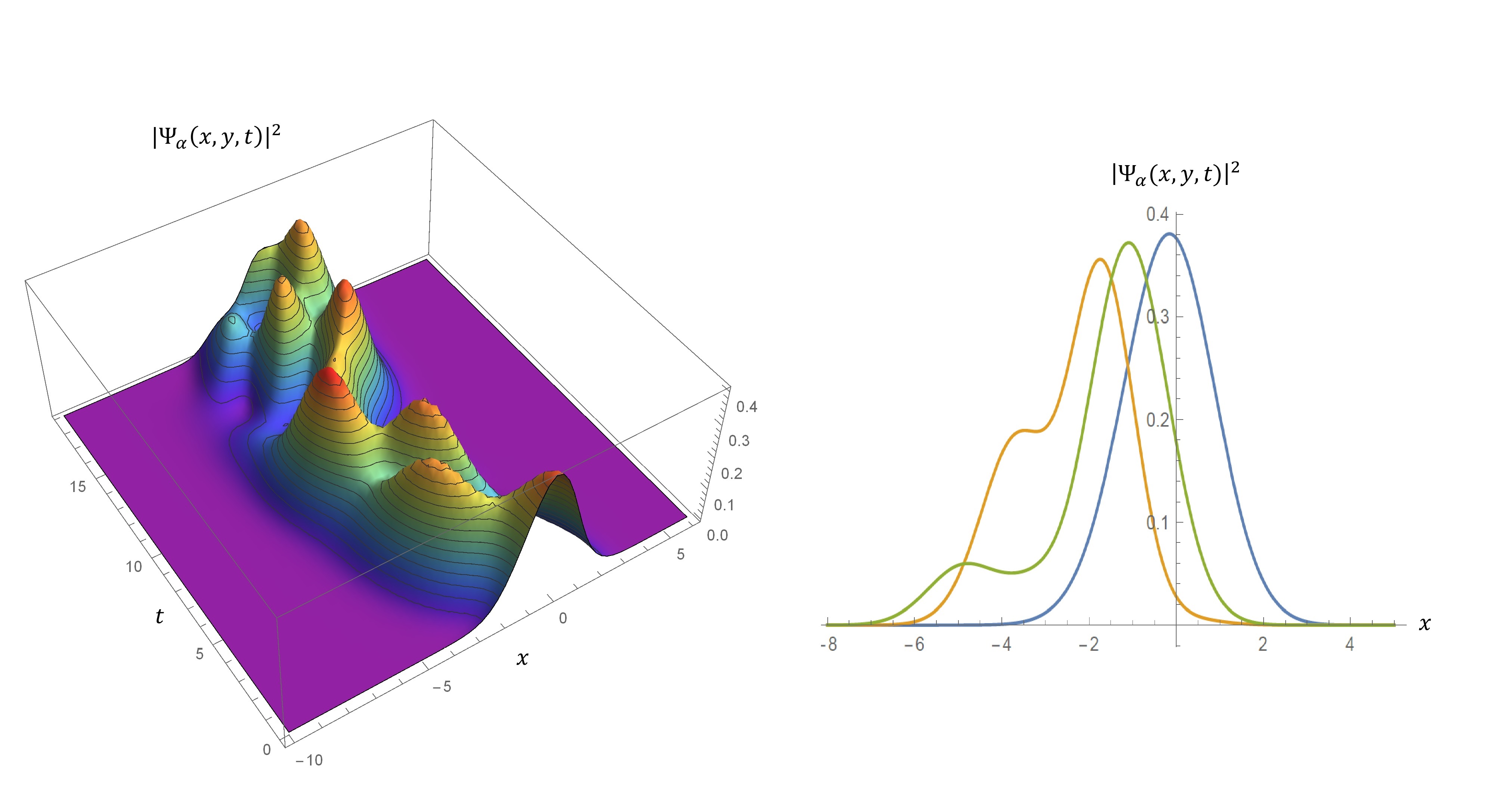}
   \caption{Left: Probability density $|\Psi_{\alpha}(x,y;t)|^{2}$ for the MGCS with $f(n)=1$ (Eq. (\ref{eq.evolmon1})), $r=1$, $\theta=0$ and $(v_{F}^{2}  \omega)^{1/2}=1$. Right: Probability density $|\Psi_{\alpha}(x,y;t)|^{2}$ for some fixed times (the suggested approximate period and some of its multiples). The blue, green and orange lines correspond to  $\tau=\left\lbrace 0, 2\pi, 4\pi\right\rbrace $, respectively.}   
  \label{fig.denmono1}
  \end{center}  
\end{figure}

\begin{figure}
\begin{center}
  \includegraphics[width=16cm]{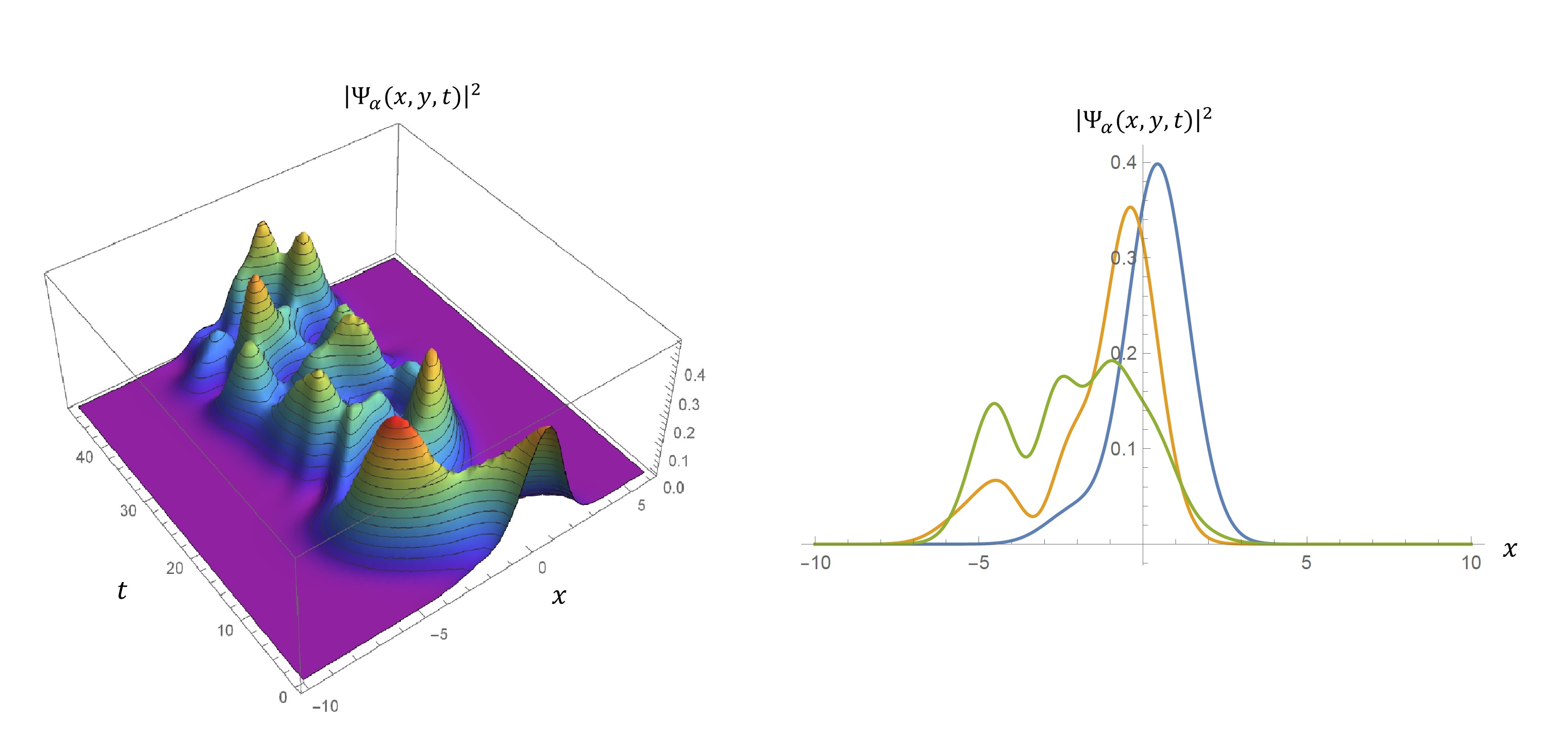}
  \caption{Left: Probability density $|\Psi_{\alpha}(x,y;t)|^{2}$ for the  MGCS with $f(n)=\sqrt{n-1}/\sqrt{n}$ (Eq. (\ref{eq.evolmon2})), $r=1$, $\theta=0$ and $(v_{F}^{2}  \omega)^{1/2}=1$. Right: Probability density $|\Psi_{\alpha}(x,y;t)|^{2}$ for some fixed times (the suggested approximate period and some of its multiples). The blue, green and orange lines correspond to $\tau=\left\lbrace 0, 5\pi, 10\pi\right\rbrace $, respectively.}  
  \label{fig.denmono2}
  \end{center}  
\end{figure}

\begin{figure}
\begin{center} 
  \includegraphics[width=16cm]{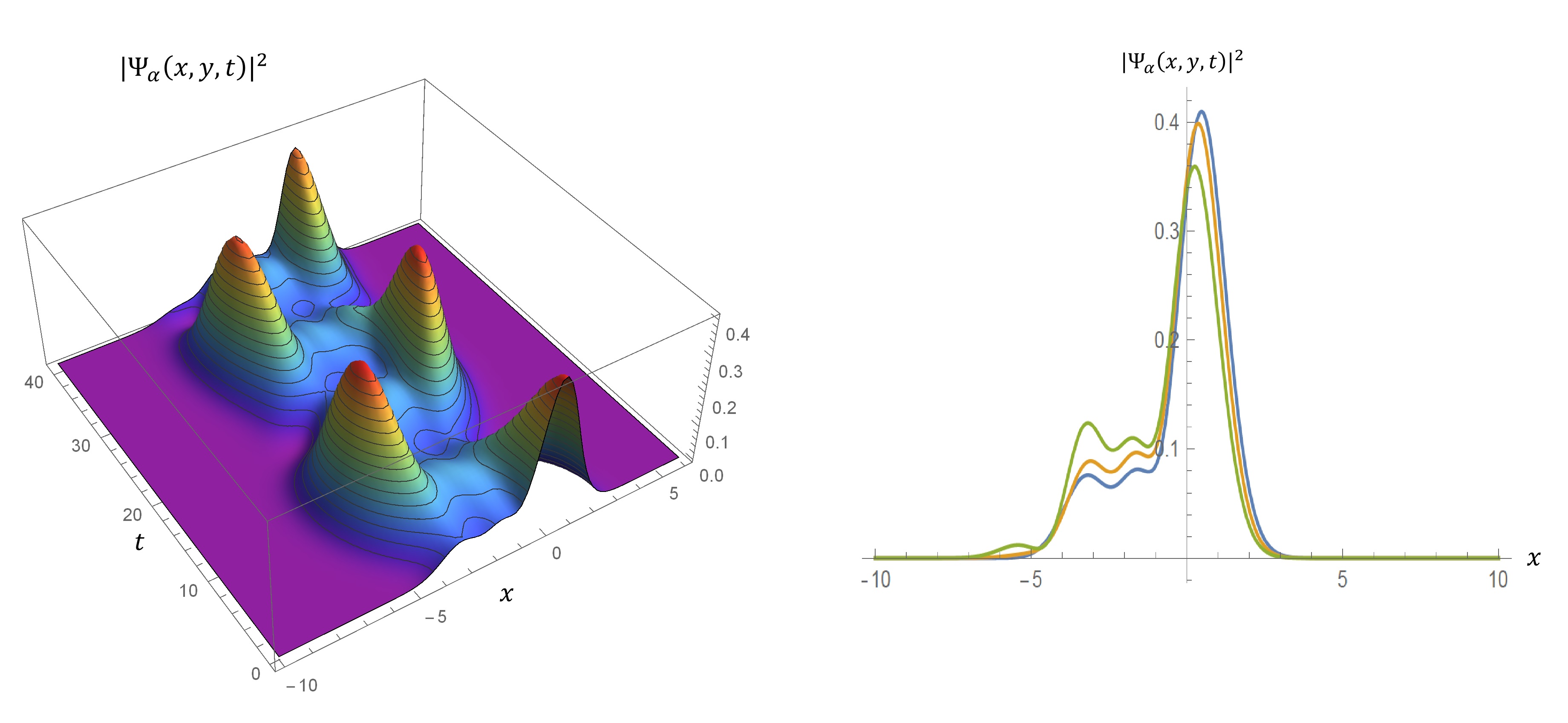}
   \caption{Left: Probability density $|\Psi_{\alpha}(x,y;t)|^{2}$ for the  MGCS with $f(n)=(n-2)\sqrt{n-1}/\sqrt{n}$ (Eq. (\ref{eq.evolmon3})), $r=1$, $\theta=0$ and $(v_{F}^{2}  \omega)^{1/2}=1$. Right: Probability density $|\Psi_{\alpha}(x,y;t)|^{2}$ for some fixed times (the suggested approximate period and some of its multiples). The blue, green and orange lines correspond to $\tau=\left\lbrace 0, 6\pi, 12\pi\right\rbrace $, respectively.}
  \label{fig.denmono3}
  \end{center}  
\end{figure}

The probability densities for the states in Eqs. (\ref{eq.evolmon1}, \ref{eq.evolmon2}, \ref{eq.evolmon3}) are shown in Figures \ref{fig.denmono1}, \ref{fig.denmono2} and \ref{fig.denmono3}, respectively.

\subsubsection{Discussion}
For monolayer graphene the energy levels $E_{n}=\hbar v_{F}\sqrt{n\omega}$ are never equally spaced, thus we cannot approximate them in general by a linear expression. Nevertheless, the graphs of $|\Psi_{\alpha}(x,y;t)|^{2}$ for the MGCS show a certain periodicity, then we can try to find an approximate period in the same way as for the BGCS (see Eq. (\ref{eq.periodo}) and the related discussion).
\\

For the states of Eq. (\ref{eq.evolmon1}) with $\vert \alpha \vert=1$, such that, $E_{0}<\langle \hat{H}\rangle_{\alpha}=0.95 \hbar v_{F}\sqrt{\omega} <E_{1}$ we obtain the possible  approximate period $\tau\simeq2\pi / v_{F} \sqrt{\omega}$. For the states of Eq. (\ref{eq.evolmon2}) we obtain that $E_{1}<\langle \hat{H}\rangle_{\alpha}=1.37 \hbar v_{F}\sqrt{\omega} <E_{2}$ and therefore a possible approximate period is $\tau\simeq5\pi / v_{F} \sqrt{\omega}$. Finally, for the states of Eq. (\ref{eq.evolmon3}) with $E_{2}<\langle \hat{H}\rangle_{\alpha}=1.53 \hbar v_{F}\sqrt{\omega}<E_{3}$ we obtain $\tau\simeq6\pi / v_{F} \sqrt{\omega}$. In Figures (\ref{fig.denmono1}-\ref{fig.denmono3}) we plot (right) the probability density for $\tau$ and some of its multiples in each case.

\section{Conclusions}
Dirac electrons in monolayer graphene interacting with magnetics fields have been studied in \cite{kuru} in terms of eigenstates and eigenvalues of the effective Hamiltonian, and more recently through coherent states \cite{erick}. Motivated by these works, in this paper we have derived as well the coherent states for electrons in bilayer graphene interacting with a constant, homogeneous magnetic field orthogonal to the graphene layers. One of the main differences (perhaps the most important one) between bilayer and monolayer graphene has to do with their energy spectrum, or Landau levels, which defines quite clearly the different time evolution they will show.
\\

We identified first the annihilation and creation operators for bilayer graphene, and then we constructed the BGCS as eigenstates of the annihilation operator with complex eigenvalue $\alpha$, which involve an arbitrary function $f$ of the number operator that can be chosen at convenience. This function leaves us a lot of freedom in the choice of the annihilation operator, and thus different sets of BGCS can be built up.
\\

Several quantities useful to study our BGCS have been calculated, the most important one being the Heisenberg uncertainty relation. For the BGCS with $f(n)=1$ the HUR has a minimum, equal to $1/2$, for $\alpha$ tending to zero while for the other two cases ($f(n)=\sqrt{n-1}/\sqrt{n}$ and $f(n)=(n-2)\sqrt{n-1}/\sqrt{n}$) this quantity reaches a maximum at the same limit, equal to $3/2$ in both cases. Let us stress that the way in which the creation and annihilation operators act on the energy eigenstates involved in the BGCS of Eq. (\ref{eq.coherente11}) almost reproduce the harmonic oscillator algebra \cite{david, oscar, hussin}. For the other two sets of BGCS (Eqs. (\ref{eq.coherente22}, \ref{eq.coherente33})) a clever choice of $f$ allowed us to exclude selectively the states with minimum energy from such expansion, which are annihilated by both operators $\hat{A}^{\pm}$ on the Hilbert space ${\cal H}$ generated by the eigenstates of the Hamiltonian in Eq. (\ref{eq.ham4}). 
\\

The probability density and probability current for the BGCS have been as well calculated for different values of $\alpha=re^{i\theta}$. We observe that as $\theta$ increases the probability density reaches a maximum which moves along $x$-direction, while the probability current shows a random behaviour, depending on the set of BGCS under consideration. We calculated also the mean energy value, which grows as the magnetic field amplitude does. For the BGCS of Eqs. (\ref{eq.coherente11}, \ref{eq.coherente22}) the behaviour of $\langle \hat{H}\rangle_{\alpha}$ is similar, unlike the BGCS of Eq. (\ref{eq.coherente33}) for which the mean energy value grows more slowly. This quantity turned out to be useful to explain the quasi-periodic behaviour seen in the time evolution of our BGCS.
\\

Specifically, the time evolution of the BGCS indicates that for linear combinations such that $n\gtrsim2$, where $n$ labels the eigenstates of the superposition, such time evolution is stable (see Figure \ref{fig.denbica3}), as for the standard coherent states \cite{Gerry}, with the same period as for the harmonic oscillator involved  ($\tau=2\pi/\omega^{*}_{c}$). This is so since the Landau levels for bilayer graphene are approximately equidistant for $n\gtrsim2$, but for BGCS where the relative contribution of the eigenstates $\Psi_{0}(x,y)$ and $\Psi_{1}(x,y)$ is significant, the time evolution turns out to be quasi-stable, as can be seen in Figures (\ref{fig.denbica1}, \ref{fig.denbica2}). 
\\

In this work we went further and calculated as well the time evolution of the MGCS derived in \cite{erick}. An important point of these states is that, since the energy spectrum goes as $\sqrt{n}$ it is not possible in general to approximate this spectrum linearly. Despite, the time evolution is approximately periodic for the three sets of coherent states
built for monolayer graphene (see Figures (\ref{fig.denmono1}, \ref{fig.denmono2}, \ref{fig.denmono3})). Thus, in this work we have proposed as well a way to calculate a possible approximate period for these states which are showing a quasi-stable motion, despite the system does not have an equidistant spectrum.
\\

Finally, let us point out that some other approaches have been implemented recently to address these kind of two-dimensional systems. For example, in \cite{wigner} the coherent states were derived for uniaxially strained graphene with non-equidistant Landau levels, and the corresponding Wigner functions (WF) were evaluated. The time dependent WF for these coherent states show fluctuations between classical and quantum behaviour, showing as well a quasi-periodic motion when they evolve in time. On the other hand, some even and odd superpositions of MGCS have been recently addressed \cite{MA20}. In our opinion, all these examples indicate a path to follow in the future, when the coherent states approach could be applied to the so-called two-dimensional Dirac materials.

\bibliographystyle{plain}

\end{document}